\documentclass[10pt,twocolumn]{IEEEtran}
\usepackage{amsmath,amscd,amssymb,amsgen,amsfonts,amsbsy}
\usepackage{mathrsfs}
\usepackage[utf8x]{inputenc}
\usepackage{color}
\usepackage{textcomp}
\usepackage{float}
\usepackage{latexsym,graphicx}
\usepackage{tikz}
\usetikzlibrary{automata,arrows,positioning,calc}
\usepackage{url}

\newcommand{\prob}[1]{\mathsf{Pr}\left( #1 \right)}

\newcommand{\remove}[1]{}

\newcommand{\comments}[1]{}

\newcommand{\qed}{\hfill $\square$}

\newtheorem{lemma}{Lemma}

\newtheorem{property}{Property}
\newtheorem{theorem}{Theorem}
\newtheorem{remark}{Remark}
\newtheorem{definition}{Definition}
\newtheorem{conjecture}{Conjecture}
\makeatother

\title{On the Whittle Index for Restless Multi-armed Hidden Markov
  Bandits}

\author{
  \begin{tabular}{cc}
    Rahul Meshram and D. Manjunath & Aditya Gopalan \\
    Deptt. of Elecl. Engg.         & Deptt. of Elecl. Commun. Engg. \\
    IIT Bombay, Mumbai INDIA       & Indian Inst. of Science, Bangalore INDIA.     
  \end{tabular}
}

\begin{document}

\maketitle
\begin{abstract}
  We consider a restless multi-armed bandit in which each arm can be
  in one of two states. When an arm is sampled, the state of the arm is
  not available to the sampler. Instead, a binary signal with a known
  randomness that depends on the state of the arm is available.
  No signal is available if the arm is not sampled. An arm-dependent
  reward is accrued from each sampling. In each time step, each arm
  changes state according to known transition probabilities which in
  turn depend on whether the arm is sampled or not sampled. Since the
  state of the arm is never visible and has to be inferred from the
  current belief and a possible binary signal, we call this the hidden
  Markov bandit. Our interest is in a policy to select the arm(s) in
  each time step that maximizes the infinite horizon discounted
  reward. Specifically, we seek the use of Whittle's index in
  selecting the arms.
  
  We first analyze the single-armed bandit and show that in general,
  it admits an approximate threshold-type optimal policy when there is
  a positive reward for the `no-sample' action. We also identify
  several special cases for which the threshold policy is indeed the
  optimal policy. Next, we show that such a single-armed bandit also
  satisfies an approximate-indexability property. For the case when
  the single-armed bandit admits a threshold-type optimal policy, we
  perform the calculation of the Whittle index for each arm. Numerical
  examples illustrate the analytical results.

\end{abstract}

\section{Introduction}
\label{sec:intro}

Restless multi-armed bandit problems are a generalization of the
classical multi-armed bandit (MAB) problem. In the MAB, the sampler
chooses one of $N$ arms in each time-step and receives a reward. Each
arm can be in one of $M$ states and the reward is dependent on the
state of the arm. The sampled arm changes state according to a known
law while the other arms are frozen. In the RMAB, all the arms change
their state at each time-step, i.e., the arms are restless. The law
that governs the change of state could depend on whether the arm was
sampled or not sampled. In this paper we introduce a class of RMAB
problems where the player never gets to observe the state of the
arm. The objective in both MAB and RMAB is to choose the sequence of
arms to sample so as to maximize a long term reward function.  We
begin with two motivating examples for the models that we introduce in
this paper.

\subsection{Motivation}
\label{sec:motivation}

Opportunistic access in time-slotted multi-channel communication
systems for Gilbert-Elliot channels \cite{Gilbert60} is being
extensively studied. In the typical model there are $N$ channels and
each channel can be in one of two states---a good state and a bad
state. Each channel independently evolves between these two states
according to a two-state Markov chain. The sender can transmit on one
of these $N$ channels in each time slot. If the selected channel is in
the good state, then the transmission is successful, and if it is in
the bad state, it is unsuccessful. The sender receives instantaneous
error-free feedback about the result of the transmission in both these
cases. If the sender knows the transition probabilities of the
channels, then using the feedback, it can calculate a `belief' for the
state of each channel in a slot. This belief may be used to select the
channel in each slot to optimize a suitable reward function. This
system and its myriad variations have been studied as restless
multi-armed bandit (RMAB) problems.

Consider a system as above except that now the probability of success
in the good state and of failure in the bad state are both less than
one and the sender knows these probabilities. This generalization of
the Gilbert-Elliot channel means that the sender does not get perfect
information about the state of the channel from the feedback. However,
it can update its a posteriori belief about the state of the channel
based on the feedback, and use this updated belief in the subsequent
slot.

As a second motivating example, consider an advertisement (ad)
placement system (APS) for a user in a web browsing session. Assume
that the APS has to place one ad from $M$ candidate ads each of which
has a known click-through probability and an expected reward
determined from the user profile. It is conceivable that the
click-through probabilities for ads in a session depend on the history
of the ads shown; users often react differently depending upon the
frequency with which an ad is shown. Some users may, due to annoyance,
respond negatively to repeated display of an ad, which has the effect
of lowering the click-through probability if they were shown this ad
in the past. Others may convert disinterest to curiosity if an ad is
repeated thereby increasing the click-through probability. Yet other
users may be more random or oblivious to what has been shown, and may
behave independently of the history.

The effect of recommendation history on a user's interest can be
modeled as follows. A state is associated with each ad and the state
changes at the end of each session (the state intuitively signifies
the interest level of the user in the ad). The transition
probabilities for this change of state depend on whether the ad is
shown or not shown to the user in the session. Assume that the state
change behavior is independent of the past and of the state change of
the other ads. Each state is associated with a value of click-through
probability and expected revenue.  The state transition and the
click-through probabilities determine the `type' or profile of the
user. In each session the APS only observes a `signal' or outcome
(click or no-click) for the ad that it displayed and no signal for
those that are not displayed. The action and the outcome is used to
update its belief about the current state of the user for each ad. The
objective of the APS would be to choose the ad in each session that
optimizes a long term objective. Clearly, this is also a RMAB with the
added generalization that the transition probabilities for the arms
depend on the action in that stage.

In this paper we analyze this generalization of the restless
multi-armed bandit---the states are never explicitly observed and the
transition probabilities depend in general on the action chosen. To
the best of our knowledge, such systems have not been considered in
the literature.

\subsection{Literature Overview}
\label{sec:literature}

Restless multi-armed bandits (RMAB) are a special class of partially
observed Markov decisions processes (POMDPs) and are in general
PSPACE-hard \cite{Papadimitriou99}, but many special cases have been
studied. An important recent application of RMABs is in dynamic
spectrum access systems, e.g.,
\cite{Nino-Mora08,Nino-Mora09,LiuZhao10,Lott10}. A common channel
shared by many heterogeneous users, each of whom see the channel as an
independent Gilbert-Elliott channel is considered in
\cite{Nino-Mora08} where an index-based policy to maximize the
discounted infinite-horizon throughput minus the transmission costs is
derived. In \cite{Nino-Mora09}, the occupancy of channels by primary
users is modeled as a two-state Markov chain. The secondary users
(SUs) sense the channel using error-prone spectrum sensors before
transmitting. Again, an index policy to maximize the infinite-horizon
discounted throughput is derived. In \cite{LiuZhao10}, the objective
is similar to that of \cite{Nino-Mora08} and it is shown that a
Whittle's index based policy is optimal. In \cite{Lott10} multiple
service classes are considered and the objective is to maximize a
utility function based on the queue occupancies. Conditions for a
myopic policy, based on instantaneous reward, to be optimal are
derived.  Myopic policies are also the subject of interest in several
other recent works, including \cite{Zhao08,Wang13,Wang14}. Utility
functions are used in \cite{Ouyang12} that considers a system similar
to that of \cite{LiuZhao10}. Opportunistic spectrum access as POMDPs
are also studied in \cite{Zhao07,Chen08,Li13}.

In much of the restless multi-armed bandit literature, including the
references in the preceding, the solution method is to seek an
`index-based' policy where the state of each arm is mapped to an index
and at each step the arms with the highest index values are played.
Whittle's index, first proposed in \cite{Whittle88}, is based on a
Lagrangian relaxation and decomposition and is a popular one; see
e.g., \cite{Veatch96,Ny08,LiuZhao10,Ouyang11,
  Avrachenkov13,Avrachenkov16}. An alternative indexing scheme is
based on partial and generalized conservation laws \cite{Nino-Mora01}
and on marginal productivity \cite{Nino-Mora09}; in this paper, we
will concentrate on the Whittle index.  The first step in determining
if an index-based policy can be used is to prove indexability. Whittle
indexability is shown by analyzing the one armed bandit as a POMDP,
the analyses of which borrows significantly from early work on POMDPs
that model machine repair problems like in
\cite{Ross71,Sernik91a,Sernik91b}.  These are described next.

In \cite{Ross71}, a machine is modeled as a two-state Markov chain
with three actions and it is shown that the optimal policy is of the
threshold type with three thresholds. In \cite{Sernik91b}, a similar
model is considered and the formulas for the optimal costs and the
policy are obtained.  This and some additional models are considered
in \cite{Sernik91a} and, once again, several structural results are
obtained. Also see \cite{Hughes80} for more such models.

The key features in the single-arm problems considered in the
preceding are as follows. One or more of the actions provides the
sampler with {\em exact} information about the state of the Markov
chain. Furthermore, the transition probability of the state of the
arms does {\em not} depend on the action. These are also the features
of each of the arms of the RMAB models discussed earlier. In this
paper we consider a model that drops both these restrictions. Since
the state is never observed but only estimated from the signals when
the arm is sampled, our model can be called a `hidden Markov restless
multi-armed bandit.'  A {\em rested} hidden Markov bandit has been
studied in \cite{Krishnamurthy01}, where the state of an arm does not
change if it is not sampled. The (arguably simpler) information
structure in a hidden rested bandit admits an analytical solution via
Gittins indices.  

A further simplification that is often made in showing indexability is
to \textit{assume,} without a formal proof, the existence of a
threshold-type optimal policy for the single-arm case, i.e., it is
optimal to play the arm if the state is higher than the threshold and
optimal to not play if the state is below the threshold as in, e.g.,
\cite{Nino-Mora08}. Under this simplification, in many cases, the
state of the arm can be mapped to an index without actually
calculating the threshold. In Section~\ref{sec:threshold} we describe
a method to do this. 
  
\subsection{Summary of the Contributions}
We now summarize the key contributions of this paper.  We consider
restless multi-armed bandits in which the transition probabilities of
the arms depends on whether the arm is played or not played. Although
the applications for this model appear to be many, to the best of our
knowledge, this is not a well-studied problem. In addition, the states of the
arms are never observed and only a belief about the state of the arm
can be computed using prior belief and the conditional probabilities
of the observation from a play of the arm. Once again, we believe such
a system has not been studied.  The preceding features make the system
hard to analyze using well known techniques. Hence we develop the
notion of an \textit{approximately threshold} type optimal policy and
prove that in general the single armed bandit that we consider admits
such an optimal policy. For some special cases of the system
parameters we also show that the single armed bandit in fact admits a
threshold-type optimal policy. We then define
\textit{approximate-indexability} and show that the arms defined by
our model also satisfy this property. This justifies the use of
Whittle's index based policy for the restless multi-armed hidden
Markov bandits. For the case when a threshold type policy is indeed
the optimal policy, we outline the procedure to compute the Whittle's
index.  Numerical examples illustrate the theory.

The model details are described in the next section.

\section{Model Description and Preliminaries}
\label{sec:model}

We consider the following restless, multi-armed bandit problem with
$N$ arms. Time is slotted and indexed by $t.$ Each arm has two states,
$0$ and $1.$ Let $X_n(t) \in \{0,1 \}$ be the state of arm $n$ at the
beginning of time $t.$ Let $A_n(t) \in \{0,1 \}$ denote the action in
slot $t$ for arm $n,$ i.e.,
\begin{eqnarray*}
  A_n(t) = \begin{cases}
    1 & \mbox{Arm $n$ is sampled in slot $t,$}\\
    0 & \mbox{Arm $n$ is not sampled in slot $t.$}
  \end{cases}
\end{eqnarray*}
We will assume that $\sum_{n=1}^N A_n(t) = 1 $ for all $t,$ exactly
one arm is sampled in each slot. Arm $n$ changes state at the end of
each slot according to transition probabilities that depend on
$A_n(t).$ Define the following transition probabilities.
\begin{eqnarray*}
  \prob{X_n(t+1)=0 | X_n(t)=0, A_n(t)=0 } & = & \lambda_{n,0}, \\
  \prob{X_n(t+1)=0 | X_n(t)=1, A_n(t)=0 } & = & \lambda_{n,1}, \\
  \prob{X_n(t+1)=0 | X_n(t)=0, A_n(t)=1 } & = & \mu_{n,0}, \\
  \prob{X_n(t+1)=0 | X_n(t)=1, A_n(t)=1 } & = & \mu_{n,1}. 
\end{eqnarray*}

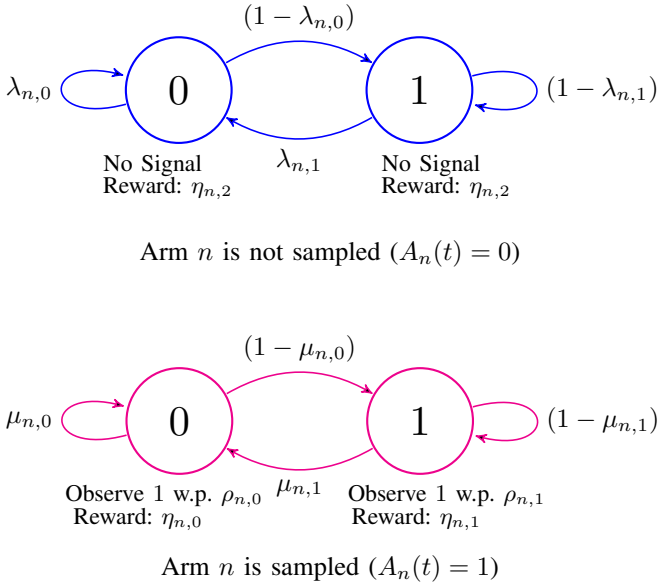
\begin{figure}

  \begin{center}
    \begin{tikzpicture}[draw=blue,>=stealth', auto, semithick, node distance=2cm]
      \tikzstyle{every state}=[fill=white,draw=blue,thick,text=black,scale=1.6]
      \node[state]    (A)                     {$0$};
      \node[state]    (B)[ right of=A]   {$1$};
      \path
      (A) edge[loop left]     node{$\lambda_{n,0}$}
      (A) edge[bend left,above,->]      node{$(1- \lambda_{n,0}) $ } (B)
      (B) edge[loop right]    node{$(1- \lambda_{n,1})$}
      (B) edge[bend left,below,->]      node{$\lambda_{n,1} $} (A); 
      \draw (-0.4,-1) node {\small{ No Signal} };
      \draw (-0.2,-1.3) node {\small{ Reward: $\eta_{n,2}$ }};
      \draw (3.3,-1) node {\small{ No Signal }};
      \draw (3.5,-1.3) node {\small{ Reward: $\eta_{n,2}$ }}; 
    \end{tikzpicture}
  \end{center}

  \vspace{2pt}
  
  \centerline{Arm $n$ is not sampled ($A_n(t) = 0$)}
  
  \vspace{7pt}
  \begin{center}
    
    \begin{tikzpicture}[draw=magenta, >=stealth', auto, semithick, node distance=2cm]
      \tikzstyle{every state}=[fill=white,draw=magenta,thick,text=black,scale=1.6]
      \node[state]    (A)                     {$0$};
      \node[state]    (B)[ right of=A]   {$1$};
      \path
      (A) edge[loop left]     node{$\mu_{n,0}$}         (A)
      edge[bend left,above,->]      node{$(1- \mu_{n,0})  $}      (B)
      (B) edge[loop right]    node{$(1- \mu_{n,1})$}     (B)
      edge[bend left,below,->]      node{$\mu_{n,1} $}         (A);    
      \draw (-0.2,-1) node {\small {Observe $1$ w.p. $\rho_{n,0}$} };
      \draw (-0.6,-1.3) node {\small{ Reward: $\eta_{n,0}$} };
      \draw (3.5,-1) node {\small{ Observe $1$ w.p. $\rho_{n,1}$} };
      \draw (3.1,-1.3) node {\small{ Reward: $\eta_{n,1}$} };
    \end{tikzpicture}
    
    \vspace{2pt} 
    
    \centerline{Arm $n$ is sampled ($A_n(t) = 1$)}
    
  \end{center}
  
  \caption{Top: State transition probabilities, the expected reward,
    and the probability of binary signal 1 being observed when the arm
    is not sampled. Bottom: The corresponding quantities when the arm
    is sampled \label{fig:Arm-MC}}
\end{figure}

In slot $t,$ if arm $n$ is in state $i$ and it is sampled, then a
binary signal $Z_n(t)$ is observed and a reward $R_{n,i}(t,1)$ is
accrued. If the arm is not sampled, then a reward $R_{n,i}(t,0)$ is
accrued and no signal is observed. Let
\begin{displaymath}
  \prob{Z_n(t)=1 \ | \ X_n(t)=i, A_n(t)=1} = \rho_{n,i}
\end{displaymath}
and denote
\begin{eqnarray*}
  R_{n,i}(t,1)=\eta_{n,i}  \hspace{0.2in} R_{n,i}(t,0) = \eta_{n,2}. 
\end{eqnarray*}
Fig.~\ref{fig:Arm-MC} illustrates the model and the parameters.

In most applications, $Z_n(t)=1$ would correspond to a `good' or
favorable output e.g., a successful transmission or click-through in
the motivating examples. Hence, we will make the reasonable assumption
that $\rho_{n,0} < \rho_{n,1}$ and $\eta_{n,0} < \eta_{n,1}$ for all
$n.$

\begin{remark}
  \begin{itemize}
  \item In the communication system example that maximizes throughput,
    no reward is accrued if there is no transmission. Also, in the APS
    example, no revenue is accrued if there is no ad displayed. Thus
    in both these cases, $\eta_{n,2} =0$ is reasonable.
  \item Further, for communication over Gilbert-Elliot channels,
    $\lambda_{n,i} = \mu_{n,i}$ for $i=0,1.$
 \end{itemize}
\end{remark}

We assume that $\lambda_{n,i},$ $\mu_{n,i},$ and $\rho_{n,i}$ are
known. The sampler cannot directly observe the state of the arm, and
hence does not know the state of the arms at the beginning of each
time slot. Instead, it can maintain the posterior or belief
distribution $\pi_n(t)$ that arm $n$ is in state $0$ given all past
actions and observations, i.e., $\pi_n(t) = \prob{X_{n}(t)=0 \; | \;
  (A_n(s), Z_n(s))_{s=1}^{t-1}}$, and is assumed known at the
beginning of slot $t.$ Thus the expected reward from sampling arm $n$
is
\begin{displaymath}
  \pi_n(t) \eta_{n,0} + (1 - \pi_n(t)) \eta_{n,1}
\end{displaymath}
and that from not sampling the arm is $\eta_{n,2}.$

Define the vector $\pi(t) = [ \pi_1, \ldots, \pi_N] \in [0,1]^N.$ Let
$H_t$ denote the history of actions and observed signals up to the
beginning of time slot $t$, i.e., $H_t \equiv (A_n(s), Z_n(s))_{1 \leq
  n \leq N, 1 \leq s < t}$. In each slot, exactly one arm is to be
sampled and let $\phi=\{\phi(t)\}_{t > 0}$ be the sampling strategy
with $\phi(t)$ defined as follows.  $\phi(t): H_t \to \{1, \ldots, N
\}$ maps the history up to time slot $t$ to the action of sampling one
of the $N$ arms at time slot $t.$ Let
\begin{displaymath}
  A^{\phi}_n(t) = 
  \begin{cases}
    1 & \mbox{if $\phi(t)=n,$} \\
    0 & \mbox{if $\phi(t) \neq n.$}
  \end{cases}
\end{displaymath}

The infinite horizon expected discounted reward under sampling policy
$\phi$ is given by
\begin{equation}
  \begin{aligned}
    && V_{\phi}(\pi) : = 
    E \left\{ \sum_{t=1}^{\infty} \beta^{t-1}  \left( 
        \sum_{n=1}^N A^{\phi}_n(t) \ ( \pi_n(t) \ \eta_{n,0} 
          \right. \right. \\
          && \left. + (1-\pi_n(t)) \ \eta_{n,1} \right) + 
        \left(1 - A^{\phi}_n(t) \right) \ \eta_{n,2} \bigg) \bigg\} .
  \end{aligned}
  \label{eq:RMAB-valfn}
\end{equation}
Here $\beta,$ $0 < \beta < 1,$ is the discount factor and the initial
belief is $\pi,$ i.e.,$\prob{X_n(1) = 0} = \pi_n.$
Our interest is in a strategy that maximizes $V_{\phi}(\pi)$ for all
$\pi \in [0,1]^N$

We begin by analyzing the single arm bandit in the next section.
Before proceeding we state the following background lemma derived from
\cite{Astrom69} that will be useful. 
The proof is given in the Appendix for the sake of completeness.
\begin{lemma}[\cite{Astrom69}]
  \label{lemma:background}
  If $f:\Re_{+}^n \to \Re_{+}$ is a convex function then for $x \in
  \Re_{+}^n,$ $g(x) := ||x||_1 f\left(\frac{x}{||x||_1}\right)$ is
  also a convex function.
\end{lemma}

{\em Notation.} For sets $A$ and $B$, $A \setminus B$ is used to
denote all the elements in $A$ which are not in $B$.

\section{Approximate Threshold Policy for the Restless 
  Single Armed Bandit with Hidden States}
\label{sec:RSAB}

For notational convenience we will drop the subscript $n$ in the
notation of the previous section. Further, we will assume that $\eta_0
= \rho_0$ and $\eta_1 = \rho_1.$ Thus $\eta_0$ and $\eta_1$ will be in
$(0,1)$ while there will be no restrictions on the range of $\eta_2.$
Extending the results to the case of arbitrary $\eta_0,$ and $\eta_1$
is straightforward.

Recall that $\pi(t) = \prob{X(t)=0 \; | \; H_t}$ and we can use Bayes'
theorem to obtain $\pi(t+1)$ from $\pi(t),$ $A(t)$ and $Z(t)$ as
follows.
\begin{enumerate}
\item If $A(t)=1,$ i.e., the arm is sampled, and $Z(t)=0$ then
  \begin{eqnarray*}
    && \pi(t+1) \ = \ \gamma_0(\pi(t)) \\
    && \hspace{10pt} := \ \frac{\pi(t) (1-\rho_0) \mu_0 + (1-\pi(t))
      (1-\rho_1) \mu_1}{\pi(t) (1-\rho_0) + (1-\pi(t)) (1-\rho_1)} . 
  \end{eqnarray*}
\item If $A(t) = 1$ and $Z(t) =1$ then
  \begin{displaymath}
    \pi(t+1) = \gamma_1(\pi(t)) := \frac{\pi(t) \rho_0 \mu_0 + 
      (1-\pi(t) )\rho_1  \mu_1}{\pi(t) \rho_0 + (1-\pi(t)) \rho_1} .
  \end{displaymath}
\item Finally, if $A(t)=0,$ i.e., the arm is not sampled at $t,$ then
  \begin{displaymath}
    \pi(t+1) = \gamma_2(\pi(t)) := \pi(t) \lambda_0 + 
    (1-\pi(t))\lambda_1 . 
  \end{displaymath}
\end{enumerate}

Recall that the policy is denoted by $\phi(t): H_t \rightarrow
\{0,1\}$ and it maps the history up to time $t$ to one of two actions
with $1$ indicating sampling the arm and $0$ indicating not sampling
the arm. The following is well known
\cite{Ross71,BertsekasV195,BertsekasV295}: (1)~$\pi(t)$ captures the
information in $H_t,$ in the sense that it is a sufficient statistic
for constructing policies depending on the history, (2)~Optimal
strategies can be restricted to stationary Markov policies, and
(3)~The optimum objective or value function, $V(\pi),$ is determined
by solving the following dynamic program
\begin{eqnarray}
  V(\pi) & = & \max \left\{ \rho(\pi) + \beta \left( \rho(\pi) 
      V(\gamma_1(\pi))  + (1-\rho(\pi)) \times 
    \right. \right. \nonumber \\
  && \hspace{0.5in} \left. \left. V(\gamma_0(\pi))\right), \ \  
    \eta_2 + \beta V(\gamma_2(\pi))\right\}, \label{eqn:V-of-pi}
\end{eqnarray}
where $\rho(\pi) = \pi \rho_0 + (1-\pi) \rho_1.$ 

Let $\pi$ be the belief at the beginning of time slot $t=1.$ Let
$V_S(\pi)$ be the optimal value of the objective function if $A(1)=1,$
i.e., if the arm is sampled, and $V_{NS}(\pi)$ be the optimal value if
$A(1) = 0,$ i.e., if the arm is not sampled. We can now write the
following.
\begin{eqnarray}
  V_S(\pi) &=& \rho(\pi) + \beta \left( \rho(\pi) V(\gamma_1(\pi)) 
  \right.  \nonumber  \\
  && \hspace{20pt} \left. + (1-\rho(\pi)) V(\gamma_0(\pi)) \right)  
  \label{eqn:VSpi}, \\
  V_{NS}(\pi) &=& \eta_2 + \beta V(\gamma_2(\pi)), \nonumber\\
  V(\pi) &=& \max\{V_S(\pi) , V_{NS}(\pi) \}.
  \label{eqn:VS,VNS,V}
\end{eqnarray}
%

% As we mention above, $\eta_2$ is to be used to index an arm and is not
% an intrinsic characteristic of the arm in the RMAB problem. 
Our first objective is to describe the structure of the value function
of the single arm system as a function of two variables---$\pi$ (the
belief) and $\eta_2$ (the reward for not sampling). We begin by
analyzing the structure of $V(\pi, \eta_2)$, $V_S(\pi,\eta_2)$, and
$V_{NS}(\pi,\eta_2)$ when one of $\pi$ or $\eta_2$ is fixed. To keep
the notation simple, when the dependence on $\eta_2$ is not made
explicit it is fixed. The following is proved in the Appendix.

\begin{lemma}
\label{lem:convexity}
 \begin{enumerate}
 \item (Convexity of value functions over the belief state) For fixed
   $\eta_2,$ $V(\pi),$ $V_{NS}(\pi)$ and $V_{S}(\pi)$ are all convex
   functions of $\pi.$
 \item (Convexity and monotonicity of value functions over passive
   reward) For a fixed $\pi,$ $V(\pi,\eta_2),$ $V_{S}(\pi,\eta_2),$
   and $V_{NS}(\pi,\eta_2)$ are non-decreasing and convex in $\eta_2.$
   \label{lemma:Vs-of-pi-rho2}
 \end{enumerate}
 \qed
\end{lemma}

We are now ready to state the first main result of this paper.
\begin{theorem}[Approximately threshold-type optimal policies] For a
  restless single-armed hidden Markov bandit with two states, $0 <
  \rho_0 < \rho_1 <1$ and a given $\eta_2,$ there exists $\beta_1 \in
  (0,1)$ such that for all $\beta \leq \beta_1$, one of the following
  statements is true.
  \begin{enumerate}
  \item
  A {\em threshold-type} optimal policy exists, i.e., there exists
  $\pi_T \in [0,1]$ for which it is optimal to sample at $\pi \in [0,
    \pi_T]$ and to not sample at $\pi \in (\pi_T,0]$.
    
  \item An {\em approximately threshold-type} optimal policy exists,
    i.e., there exist $\epsilon > 0$ and $\pi_T, \pi^\circ \in [0,1]$
    with $\rho(\pi^\circ) = \eta_2$ such that an optimal policy
    samples at $\pi \in [0, \pi_T] \setminus (\pi^\circ - \epsilon,
    \pi^\circ + \epsilon)$ and does not sample at $\pi \in (\pi_T,1]
      \setminus (\pi^\circ - \epsilon, \pi^\circ + \epsilon)$.
  \end{enumerate}
  \label{thm:single-threshold}
\end{theorem}
\begin{remark}
The result essentially states that, under a suitable
discount factor $0 < \beta < \beta_1$, an optimal policy has a
threshold-structure at all belief states $[0,1]$, except possibly
within a small neighbourhood of radius $\epsilon$ around the belief
state $\pi^\circ$.
\end{remark}

\begin{IEEEproof}
  Define the intervals $S_1$ and $S_2$ as follows.  
  \begin{eqnarray*}
    S_1 & = & \{\pi: \pi \in [0,1]: \ \ \eta_2 < \rho(\pi) \} \\
    S_2 & = & \{\pi: \pi \in [0,1]: \ \ \eta_2 \geq \rho(\pi) \}
  \end{eqnarray*}
  In the following we will use the subscript $\beta$ to make the
  dependence of $V_S,$ $V_{NS}$ and $V$ on $\beta$ explicit. 
  For notational convenience, let us define
  \begin{eqnarray*}
    V_{a, \beta}(\pi,\eta_2) & := &   \left[ \rho(\pi) 
      V_{\beta} (\gamma_1(\pi),\eta_2) \right. + \\
      && \hspace{15pt} \left.
      (1-\rho(\pi))V_{\beta}(\gamma_0(\pi),\eta_2) \right]. 
  \end{eqnarray*}
  From \eqref{eqn:VSpi}, we see that $\beta V_{a,\beta}(\pi,\eta_2)$ 
  is the second term for the expression for $V_{S,\beta}(\pi,\eta_2).$ 
  %From Lemma~\ref{lemma:policy-rL,rH}, we know that for all $\beta,$
  %$V_S(\pi)$ and $V_{NS}(\pi)$ intersect at least once in $[0,1]$
  %when $\eta_H \leq \eta_2 \leq \eta_L.$ 
%  Define
%  %
%  \begin{displaymath}
%     \pi_T(\eta_2) := \inf \ \{\pi \in [0,1]: 
%     V_{S,\beta}(\pi,\eta_2) = V_{NS,\beta}(\pi,\eta_2) \}. 
%  \end{displaymath} 
%  % 
%  Note that $\pi_T(\eta_2)$ may be a null set for some values of
%  $\eta_2.$ 
%  \revision{ 
For a fixed $\beta,$ $V_{\beta}(\pi,\eta_2)$ and $ V_{a,
  \beta}(\pi,\eta_2)$ are bounded for all $\pi \in [0,1]$; this
follows from $\rho_0, \rho_1,$ and $\eta_2$ being bounded and $0<\beta
<1.$ Further, in Appendix \ref{subsec:proof-V-increase-beta}, we show
that for fixed $\pi$ and $\eta_2,$ $V_{\beta}(\pi, \eta_2)$ is an
increasing function of $\beta.$

  For each belief state $\pi \in [0,1]$ satisfying $\eta_2 \neq
  \rho(\pi) = \pi \rho_0 + (1-\pi) \rho_1$, let us define\footnote{We
    follow the standard convention that $\sup\{x: x \in \emptyset\} =
    -\infty$ (resp. $\inf\{x: x \in \emptyset\} = +\infty$), where
    $\emptyset$ denotes the empty set, and in this case we say that
    the supremum (resp. infimum) does not exist or is not finite.}
  $\beta_1(\pi)$ as
  \begin{eqnarray}
    \beta_1(\pi) & := & \sup \bigg\{ \beta \in (0,1) \ : \nonumber \\
    && \hspace{4pt} \frac{|\eta_2 -\rho(\pi)|}{\beta}
      \ >  \ \left|V_{\beta} (\gamma_2(\pi)) - V_{a, \beta}(\pi)\right| \bigg\}. 
      \label{eq:beta_1}
  \end{eqnarray}
  Such a $\beta_1(\pi)$ exists in $(0,1]$ because, as we have argued
  previously, the difference between $V$ and $V_a$ is bounded, and
  moreover, $|\eta_2 -\rho(\pi)| > 0$. Now define, for any $\epsilon \geq  0$, the set
  \[ C_\epsilon := \left\{ \pi \in [0,1]: |\rho(\pi) - \eta_2| \geq
    \epsilon \right\},\] and the quantity 
  \begin{displaymath}
    \beta_{1,\epsilon} := \inf \left\{ \beta_1(\pi): \pi \in C_\epsilon \right\}.
  \end{displaymath}

  It follows that $\beta_{1,\epsilon}$ is finite (i.e., the set
  $C_\epsilon$ is nonempty) whenever either 

  \begin{enumerate}
  \item $\eta_2 \notin \{\rho(\pi): \pi \in [0,1]\}$. In this case we
    will have a (perfect) threshold-type optimal policy by taking
    $\epsilon = 0$ $\Rightarrow$ $C_\epsilon = [0,1]$ as will follow
    below.

  \item $\eta_2 \in \{\rho(\pi): \pi \in [0,1]\}$ and $\epsilon <
    \max\{\pi^\circ, 1-\pi^\circ\}$ with $\rho(\pi^\circ) =
    \eta_2$. Note that in this case, $S_1 = [0, \pi^\circ)$ and $S_2 =
      [\pi^\circ, 1]$. Here, by taking any $0 < \epsilon <
      \max\{\pi^\circ, 1-\pi^\circ\}$, we will have an approximate
      threshold-type optimal policy as will follow below.  We remark
      that in this case, for any $\epsilon$ as above, it can be argued
      that $\beta_{1,\epsilon}$ is positive as follows. Given the
      expected reward parameters $\rho_0$, $\rho_1$ and $\eta_2$, let
      $u := \max\{\rho_0, \rho_1, \eta\}$, so that $|V_\beta(\cdot)|
      \leq \frac{u}{1-\beta}$ uniformly, implying that
      $\left|V_{\beta} (\gamma_2(\pi)) - V_{a, \beta}(\pi)\right| \leq
      \frac{2u}{1-\beta}$ for all $\pi$. Now, for any $\pi \in
      C_\epsilon$, we have
      \begin{align*}
      	&\delta := \frac{\epsilon}{2u + \epsilon} \\
      	\Rightarrow \; &\frac{2u}{1-\delta} = \frac{\epsilon}{\delta} \leq \frac{|\rho(\pi) -\eta_2|}{\delta} \\
      	\Rightarrow \; &\left|V_{\delta} (\gamma_2(\pi)) - V_{a, \delta}(\pi)\right| \leq \frac{|\rho(\pi) -\eta_2|}{\delta} \\
      	\Rightarrow \; &\delta \in \beta_1(\pi),
      \end{align*}
      and so the infimum of all such numbers must satisfy $\beta_{1,\epsilon} \geq \delta = \frac{\epsilon}{2u + \epsilon} > 0$.
        \\
  \end{enumerate}

  We now claim that for any $\epsilon$ for which $\beta_{1,\epsilon}$
  is finite, and for any $\beta < \beta_{1,\epsilon}$, the optimal
  policy chooses to sample whenever the belief state is in the region
  $S_1 \cap C_\epsilon$, and to not sample in the region $S_2 \cap
  C_\epsilon$.

  First, for $\pi \in S_1 \cap C_\epsilon$, $V_{S,\beta}(\pi,\eta_2) >
  V_{NS,\beta}(\pi,\eta_2).$ To see this, write
  \begin{eqnarray*}
    V_{S,\beta}(\pi,\eta_2) - V_{NS,\beta}(\pi,\eta_2) & =  & 
    (\rho(\pi) -\eta_2) \\
    && \hspace{-30pt} -\beta\left(
      V_{\beta} (\gamma_2(\pi),\eta_2) - V_{a, \beta}(\pi,\eta_2) \right) . 
  \end{eqnarray*}    
  For $\pi \in S_1,$ the term in the first parentheses in the right
  hand side (RHS) above is positive. We now consider two cases. If the
  term in the second parentheses is negative, then the RHS is positive
  and the claim holds.  On the other hand, if the term is positive,
  then from the definition of $\beta_{1,\epsilon},$ for all $\beta <
  \beta_1,$ the second term is less than the first and for this case
  too the claim follows.
  
  On the other hand, for $\pi \in S_2 \cap C_\epsilon$, the claim
  follows by observing that
  \begin{displaymath}
     V_{a, \beta}(\pi,\eta_2)  - V_{\beta}(\gamma_2(\pi),\eta_2) < 
     \frac{\eta_2 - \rho(\pi)}{\beta}.
  \end{displaymath}
  whenever $\beta < \beta_1(\pi).$ Hence $V_{S}(\pi) < V_{NS}(\pi)$
  for $\beta < \beta_{1,\epsilon}$. This completes the proof.
\end{IEEEproof}

This theorem states that if $\eta_2 \in [\rho_0, \rho_1],$ then there
is at least an approximate threshold policy. Of course if $\eta_2 <
\rho_0,$ then the policy is to always sample corresponding to a
threshold policy with $\pi_T =1.$ Similarly, $\eta_2 > \rho_1$
corresponds to a threshold policy with $\pi_T =0. $ 
    
\subsection{Special case: Existence of a threshold-type optimal policy}

In Theorem~\ref{thm:single-threshold}, we have introduced two
approximations---a restriction on the range of $\beta,$ and also a
`hole' in the range of $\pi,$ the state of the arm, for which we do
not know the optimal policy. We now consider a special case where we
do not need to use these approximations, i.e., the optimal policy is
always of the threshold type.  The key idea behind these is to use
Lemma 2 and Lemma 3 (below) and argue that the difference between the
value functions from sampling and not sampling,
$(V_S(\pi)-V_{NS}(\pi)),$ which we call the \emph{sampling advantage,}
is monotonic in $\pi$ under these special cases of $\lambda$s and
$\mu$s.

Assume $\eta_0 = \rho_0$ and $\eta_1 = \rho_1.$ We will need the
following lemma that shows that for a suitable range of parameter
values, $(V_S(\pi) - V_{NS}(\pi))$ is monotonic.
\begin{lemma}
  \textit{(Monotonicity of the sampling advantage)} For a fixed
  $\eta_2$ and $\beta \in (0,1]$, $(V_S(\pi) - V_{NS}(\pi))$ is a
  decreasing function in $\pi$ for the following cases.
  \begin{enumerate}
  \item $0 \leq \mu_0 - \mu_1 \leq \frac{1}{5}$ and $\vert \lambda_0 -
    \lambda_1 \vert \leq \frac{1}{5}.$
  \item $0 \leq \mu_1 - \mu_0 \leq \frac{1}{3}$
    $\vert \lambda_0 - \lambda_1 \vert \leq \frac{1}{3}.$
  \end{enumerate}
  \label{lemma:diff-value-S-NS-monotone}
\end{lemma}
The proof is provided in the Appendix. This now enables us to state
the following result.
\begin{theorem}[Exact threshold-type optimal policies] 
  For a restless single-armed
  hidden Markov bandit with two states, $0 < \rho_0 =\eta_0 <
  \rho_1=\eta_1 <1$ and given $\eta_2,$ for all $\beta \in (0,1],$ a
    {\em threshold-type} optimal policy exists, i.e., there exists
    $\pi_T \in [0,1]$ for which it is optimal to sample at $\pi \in
    [0, \pi_T]$ and to not sample at $\pi \in (\pi_T,0]$, whenever
      \begin{enumerate}
      \item $0 \leq \mu_0 - \mu_1 \leq \frac{1}{5}$ and 
        $\vert \lambda_0 - \lambda_1 \vert \leq \frac{1}{5},$ or 
      \item $0 \leq \mu_1 - \mu_0 \leq \frac{1}{3}$ and 
        $\vert \lambda_0 - \lambda_1 \vert \leq \frac{1}{3}.$
      \end{enumerate}
      \label{thm:special-case-single-threshold} 
\end{theorem}

\begin{IEEEproof}
  For a fixed $\beta$ and $\eta_2$, from
  Lemma~\ref{lemma:diff-value-S-NS-monotone}, we also know that
  $(V_S(\pi) - V_{NS}(\pi))$ is decreasing in $\pi.$ Also $V_S(\pi)$
  and $V_{NS}(\pi)$ are convex in $\pi.$ This implies that there is at
  most one point in $(0,1)$ at which $V_S(\pi)$ and $V_{NS}(\pi)$
  intersect. This completes the proof.
\end{IEEEproof}
\begin{remark}
  Note that we do not make any assumption on the ordering of
  $\lambda_0$ and $\lambda_1$ except that the absolute difference is
  bounded by $\frac{1}{5}$ or by $\frac{1}{3}$ which in turn depends
  on the ordering of $\mu_0$ and $\mu_1.$
\end{remark}

\subsection{Numerical Examples}
\label{sec:rsab-numericals}

Theorem~\ref{thm:single-threshold} introduces two approximations---an
upper bound on the discount factor, and a `hole' in $[0,1]$ where we
do not know the optimal policy. We believe that this is just an
artifact of the proof technique and that the restriction on $\beta$
and hole need not actually exist. To see this we conducted an
extensive numerical experiments in which the value functions were
evaluated numerically using value iteration.
Fig.~\ref{plots:VS-VNS-set1} in the Appendix shows the plots for
$V_S(\pi)$ and $V_{NS}(\pi)$ for a sample set of $\mu_i,$ $\lambda_i,$
and $\rho_i$ for different values of the discount factor $\beta$ and
$\eta_2.$ All our results indicated that there is just one threshold
even when $\beta$ is very large and even close to 1. This leads us to
believe that both the approximations may not be needed, and to state
the following conjecture.

\begin{conjecture}[Existence of threshold-type optimal policies] For
  a restless single-armed hidden Markov bandit with two states with
  $0 < \rho_0 < \rho_1 <1$, a {threshold-type} optimal policy exists,
  i.e., there exists $\pi_T \in [0,1]$ for which it is optimal to
  sample at $\pi \in [0, \pi_T]$ and to not sample at
  $\pi \in (\pi_T,0]$.
  \label{conj:threshold}
\end{conjecture}

\section{Approximate Indexability of the Restless 
  Multi-armed Bandit with Hidden States}
\label{sec:rmab}

We are now ready to analyze the general case of the
\textit{multi-armed} bandit setting. As we have discussed in the
introduction, finding the optimal policy is, in general, a hard
problem. A heuristic that is widely used in optimally selecting the
arm at each time step is due to Whittle \cite{Whittle88}. This
heuristic is in general suboptimal but has a good empirical
performance and a large class of practical problems use this policy
because of its simplicity. In some cases, it can also be shown to be
optimal, e.g., \cite{LiuZhao10}. The arm selection in each time slot
proceeds as follows. The belief vector $\pi(t)$ is used to calculate
the Whittle's index (defined below) for each arm and the arm with the
highest index is sampled. To be able to compute such an index for each
arm, we first need to determine if the arm is indexable. Toward
determining indexability, let us first define,
\begin{eqnarray*}
  \mathcal{P}_\beta(\eta_2) &:=& \{ \pi \in [0,1]: V_{S,\beta}(\pi,\eta_2) \leq V_{NS,\beta}(\pi,\eta_2) \}. 
\end{eqnarray*}
In other words, for a given $\beta,$ $\mathcal{P}_\beta(\eta_2)$ is
the set of all belief states $\pi$ for which not sampling is the
optimal action. From \cite{Whittle88}, indexability of an arm is
defined as follows.
\begin{definition}[Indexability]
 An arm in the single-armed bandit process is indexable if 
 $\mathcal{P}_\beta(\eta_2)$ monotonically
  increases from $\emptyset$ to the entire state space $[0,1]$ as
  $\eta_2$ increases from $-\infty$ to $\infty$, i.e.,
  $\mathcal{P}_{\beta}(\eta_2^{(a)}) \setminus
  \mathcal{P}_{\beta}(\eta_2^{(b)}) = \emptyset$ whenever
  $\eta_2^{(a)} \leq \eta_2^{(b)}$. A restless multi-armed bandit
  problem is indexable if every arm is indexable.
  \label{def:indexable}
\end{definition}
\begin{definition}[Approximate or $\epsilon$-indexability]
  For $\epsilon \geq 0$, an arm is said to be {\em $\epsilon$-indexable}
  for the single-armed bandit process if, for $\eta_2^{(a)} <
  \eta_2^{(b)}$, we have $\mathcal{P}_{\beta}(\eta_2^{(a)}) \setminus
  \mathcal{P}_{\beta}(\eta_2^{(b)}) \subseteq [\tilde{\pi} - \epsilon,
  \tilde{\pi} + \epsilon]$ for some $\tilde{\pi} \in [0,1]$.
  \label{def:approxindexable}
\end{definition}

Next we define the Whittle index for an arm in state $\pi.$ 
\begin{definition}
  If an indexable arm is in state $\pi,$ its Whittle index 
  $W(\pi)$ is
  \begin{eqnarray}
    W(\pi)  &=& \inf \{\eta_2 \in \mathbb{R}: V_{S,\beta}(\pi,\eta_2) =
    V_{NS,\beta}(\pi,\eta_2) \}.
    \label{eqn:Whittle-index}
  \end{eqnarray}
\label{def:whittleind}
\end{definition}

In other words, $W(\pi)$ is the minimum value of the {\em no-sampling
  subsidy} $\eta_2$ such that the optimal action at belief state $\pi$
is to not sample an arm. Our next objective is to show that the arms
in our problem are all indexable. Showing indexability, at a high
level, requires us to show that the set $\mathcal{P}_\beta(\eta_2)$
increases monotonically as $\eta_2$ increases.  We now prove the
second key result of the paper, on the approximate-indexability of an
arm.
\begin{theorem} ($\epsilon$-Indexability of the single-armed bandit) For a
  restless single-armed hidden Markov bandit with two states, $0 <
  \rho_0 < \rho_1 <1$, there exists a $\beta_2,$ $0 < \beta_2 < 1,$
  and $\epsilon \geq 0$ such that for all $\beta < \beta_2$, the arm
  is $\epsilon$-indexable.
  \label{thm:indexable}
\end{theorem}

\begin{IEEEproof}
  First, we make the intuitive claim that there exist finite $\eta_L$,
  $\eta_H$, such that $\mathcal{P}_\beta(\eta_2) = \emptyset$
  (resp. $\mathcal{P}_\beta(\eta_2) = [0,1]$) when $\eta_2$ is less
  than (resp. greater than) $\eta_L$ (resp. $\eta_H$). This is because
  the rewards are finite and the objective function is a discounted
  reward.

%First, observe that there exist finite $\eta_2^-$, $\eta_2^+$, such that $\mathcal{P}_\beta(\eta_2) = \emptyset$ (resp. $\mathcal{P}_\beta(\eta_2) = [0,1]$) when $\eta_2$ is less than (resp. greater than) $\eta_2^-$ (resp. $\eta_2^+$). For instance, one can take $\eta_2^- = \min\{\rho_0, \rho_1\}$, $\eta_2^+ = \max\{\rho_0, \rho_1\}$. Define $\eta_L = \sup\{\eta_2: \mathcal{P}_\beta(\eta_2) = \emptyset \}$ and $\eta_H = \inf \{\eta_2: \mathcal{P}_\beta(\eta_2) = [0,1] \}$ to be the largest and smallest such limits.
%Also, it is not hard to see that $\mathcal{P}_\beta(\eta_2) = \emptyset$ (resp. $\mathcal{P}_\beta(\eta_2) = [0,1]$) whenever $\eta_2 < \eta_L$ (resp. $\eta_2 > \eta_H$). 
%\AG{What would be a quick proof of the claim above?}
%Now let us define, for $\eta_2 \in [\eta_L, \eta_H]$, the quantity $\pi_T(\eta_2) = \inf\{0 \leq \pi \leq 1: V_S(\pi,\eta_2) = V_{NS}(\pi, \eta_2) \}$. 
  
\begin{lemma}
\label{lem:grad-VS-less-grad-VNS1}
If for each $\eta_2 \in [\eta_L, \eta_H]$,
  \begin{equation}
    \frac{\partial V_S(\pi,\eta_2)}{\partial \eta_2} \bigg 
    \rvert_{\pi = \pi_T(\eta_2)} \ < \
    \frac{\partial V_{NS}(\pi,\eta_2)}{\partial \eta_2} \bigg 
    \rvert_{\pi=\pi_T(\eta_2)},
    \label{eq:grad-VS-less-grad-VNS1}
  \end{equation}
  then $\pi_T(\eta_2)$ is a monotonically decreasing function of
  $\eta_2$ in $[\eta_L, \eta_H]$. Here, $\frac{\partial V_S(\pi,\eta_2)}{\partial \eta_2}$ denotes the right partial derivative of $V_S(\pi, \cdot)$. 
  \label{lemma:indexability}
\end{lemma}
% 
%Thus showing that \eqref{eq:grad-VS-less-grad-VNS1} holds for all
%$\eta_2 \in [\eta_L, \eta_H]$ implies indexability.

Henceforth, we assume that $\eta_2 \in [\eta_L, \eta_H]$.

  Taking the partial derivative of $V_{S,\beta}(\pi,\eta_2),$ and
  $V_{NS,\beta}(\pi,\eta_2)$ with respect to $\eta_2$ we obtain
  \begin{eqnarray}
    \frac{\partial V_{S,\beta}(\pi,\eta_2)}{\partial \eta_2}  & = & \beta \left[\rho(\pi)
     \frac{\partial     V_{\beta}\left(\gamma_1(\pi),\eta_2\right)}{\partial \eta_2} \right. 
     \nonumber \\
    && \hspace{5pt} \left.+ (1-\rho(\pi)) \frac{\partial V_{\beta}
    \left(\gamma_0(\pi),\eta_2\right)}{\partial \eta_2} \right],
    \label{eq:diff_VS_rho2} \label{eq:diff_VS_r2}\\
    \frac{\partial V_{NS,\beta}(\pi,\eta_2)}{\partial \eta_2} & = & 1+ \beta 
    \frac{\partial V_{\beta}\left(\gamma_2(\pi),\eta_2\right)}{\partial \eta_2}.
    \label{eq:diff_VNS_r2} 
  \end{eqnarray}

Taking \eqref{eq:diff_VNS_r2} - \eqref{eq:diff_VS_r2}, we obtain
\begin{eqnarray*}
  &&  \frac{\partial V_{NS,\beta}(\pi,\eta_2)}{\partial \eta_2}  - \frac{\partial V_{S,\beta}(\pi,\eta_2)}{\partial \eta_2} = \\
  && 1+ \beta \frac{\partial V_{\beta}\left(\gamma_2(\pi),\eta_2\right)}
  {\partial \eta_2}  - \ \beta \left[r(\pi) \frac{\partial V_{\beta}\left(\gamma_1(\pi),\eta_2\right)}
    {\partial \eta_2} \right. \\
  && \hspace{15pt} \left.  +(1-r(\pi)) \frac{\partial V_{\beta}
      \left(\gamma_0(\pi),\eta_2\right)}{\partial \eta_2} \right] . 
\end{eqnarray*}
We now show that the above is greater than 0 at $\pi = \pi_T(\eta_2)$.
After rearranging the terms this requirement reduces to requiring that
\begin{eqnarray}
  \nonumber
 && \frac{1}{\beta} > \Bigg \{   
    \left[\rho(\pi) \frac{\partial V_{\beta}\left(\gamma_1(\pi),\eta_2\right)}
    {\partial \eta_2}   \right. \nonumber \\
   && \hspace{15pt} \left. \left. + (1-\rho(\pi)) \frac{ \partial V_{\beta}\left(\gamma_0(\pi),\eta_2\right)}
    {\partial \eta_2} \right]_{\pi = \pi_T(\eta_2)} \right. \nonumber \\
    && \hspace{25pt} - \bigg [ \frac{\partial V_{\beta}\left(\gamma_2(\pi),\eta_2\right)}{\partial \eta_2}
    \bigg ]_{\pi = \pi_T(\eta_2)} \Bigg \} . 
  \label{eq:beta_2}
\end{eqnarray}

Since $V_{\beta}(\pi,\eta_2)$ is a bounded function for fixed $\beta,$
$0<\beta<1,$ finite $\eta_2,$ and $\pi \in[0,1],$ the partial (right)
derivative of $V_{\beta}(\pi,\eta_2)$ with respect to $\eta_2$ is also
bounded. This means that we can find $\beta_2$ such that for all $0 <
\beta < \beta_2,$ the conclusion \eqref{eq:grad-VS-less-grad-VNS1} of
Lemma \ref{lem:grad-VS-less-grad-VNS1} holds.  We will also require
$\beta$ to be in $(0,\beta_1)$ with $\beta_1$ from the conclusion of
Theorem \ref{thm:single-threshold}.

Thus, letting $\beta_3 = \min\{\beta_1,\beta_2 \},$ we get that the
first crossing point $\pi_T(\eta_2)$ is monotone non-decreasing with
$\eta_2$.

To complete the proof, note that the only other states $\pi >
\pi_T(\eta_2)$ at which the optimal action {\em may} play the
no-sampling action must lie within an $\epsilon$-radius hole around
$\pi^\circ$, as shown in Theorem \ref{thm:single-threshold}. This
establishes the conclusion of the theorem.
\end{IEEEproof}

Under the conditions of
Theorem~\ref{thm:special-case-single-threshold}, we can do away with
the approximations of Theorem~\ref{thm:indexable} and explicitly
characterize a bound on the discount $\beta$ required for
indexability. Specifically, we state the following.
\begin{theorem}
  For a restless single-armed hidden Markov bandit with two states, $0
  < \rho_0<\rho_1<1,$ and finite $\eta_2$ if either
  \begin{enumerate}
  \item $0 \leq \mu_0 - \mu_1 \leq \frac{1}{5}$ and 
    $\vert \lambda_0 - \lambda_1 \vert \leq \frac{1}{5},$ or
  \item $0 \leq \mu_1 - \mu_0 \leq \frac{1}{3}$ and 
    $\vert \lambda_0 - \lambda_1 \vert \leq \frac{1}{3}$
  \end{enumerate} 
   is true then for all $\beta \in (0, 1/3),$ the arm is indexable.
\label{thm:spcial-case-indexable}
\end{theorem}
\begin{IEEEproof}
  We know from Theorem~\ref{thm:special-case-single-threshold} that
  the optimal policies are threshold type with single threshold, i.e.,
  $\pi_T(\eta_2)$ is unique for given $\eta_2.$ Further, we can obtain
  the following inequalities using induction techniques as in, for
  example, Lemma~\ref{lemma:Vs-of-pi-rho2}
  \begin{eqnarray}
    \bigg \lvert \frac{\partial V(\pi, \eta_2)}{\partial \eta_2} \bigg \rvert , 
    \bigg \lvert \frac{\partial V_S(\pi, \eta_2)}{\partial \eta_2} \bigg \rvert , 
    \bigg \lvert \frac{\partial V_{NS}(\pi, \eta_2)}{\partial \eta_2} \bigg \rvert
     \  \leq \ \frac{1}{1-\beta} 
    \label{eqn:value-eta2-diff-ineq}
  \end{eqnarray}
  % Following the arguments in Theorem~\ref{thm:indexable}, we need to
  show that \eqref{eq:beta_2} is true for range of the parameters that
  we consider here.  This is done by using
  \eqref{eqn:value-eta2-diff-ineq}, and upper bounding the RHS of
  \eqref{eq:beta_2} as follows.
  \begin{eqnarray*}
    RHS &\leq & \rho(\pi) \frac{1}{1-\beta} +
    (1-\rho(\pi))\frac{1}{1-\beta} + \frac{1}{1-\beta}, \\ 
    &=& \frac{2}{1-\beta}.
  \end{eqnarray*}
  If $\beta < 1/3,$ then $ \frac{2}{1-\beta} <  \frac{1}{\beta}$ implying 
  \eqref{eq:beta_2} to complete the proof. 
\end{IEEEproof}

\begin{remark}
  Theorem~\ref{thm:indexable} tells us that the restless multi-armed
  bandit with hidden states is approximately indexable. Like in
  Theorem~\ref{thm:single-threshold}, we believe that the
  approximation is just an artifact of the proof technique and result
  is possibly more generally true and also without the restriction on
  $\beta.$ This is also borne out by extensive numerical study that we
  conducted.  In Fig.~\ref{plots:pi_T} we show a sample plot of
  $\pi_T(\eta_2),$ the threshold belief as a function of the passive
  subsidy $\eta_2$ for different $\beta.$ We see that $\pi_T$
  increases with $\eta_2$ leading us to believe that indexability is
  more generally true.
\begin{figure*}
  \begin{center}    
    \begin{tabular}{cc}
    \includegraphics[width=0.8\columnwidth]{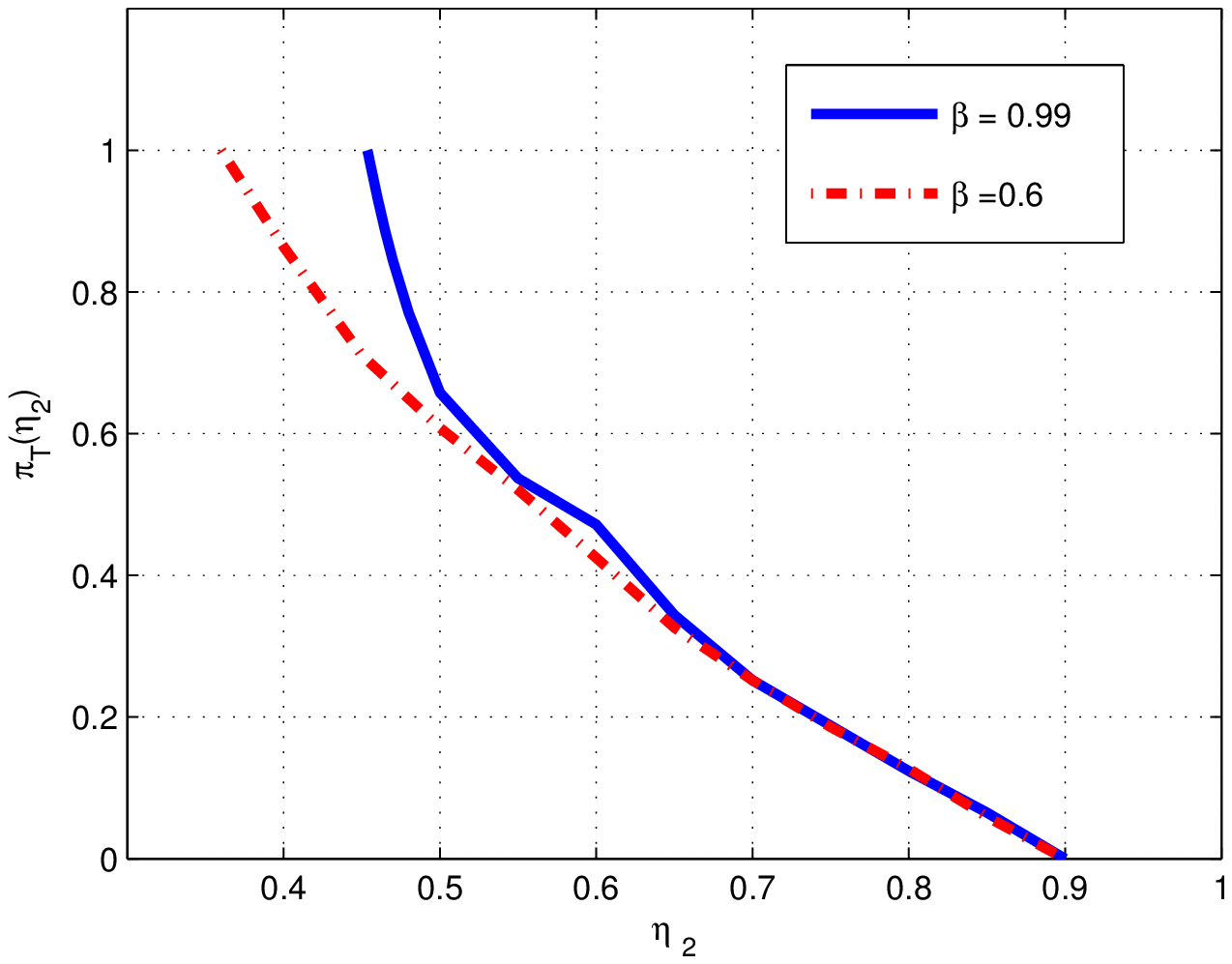}
    &
    \includegraphics[width=0.8\columnwidth]{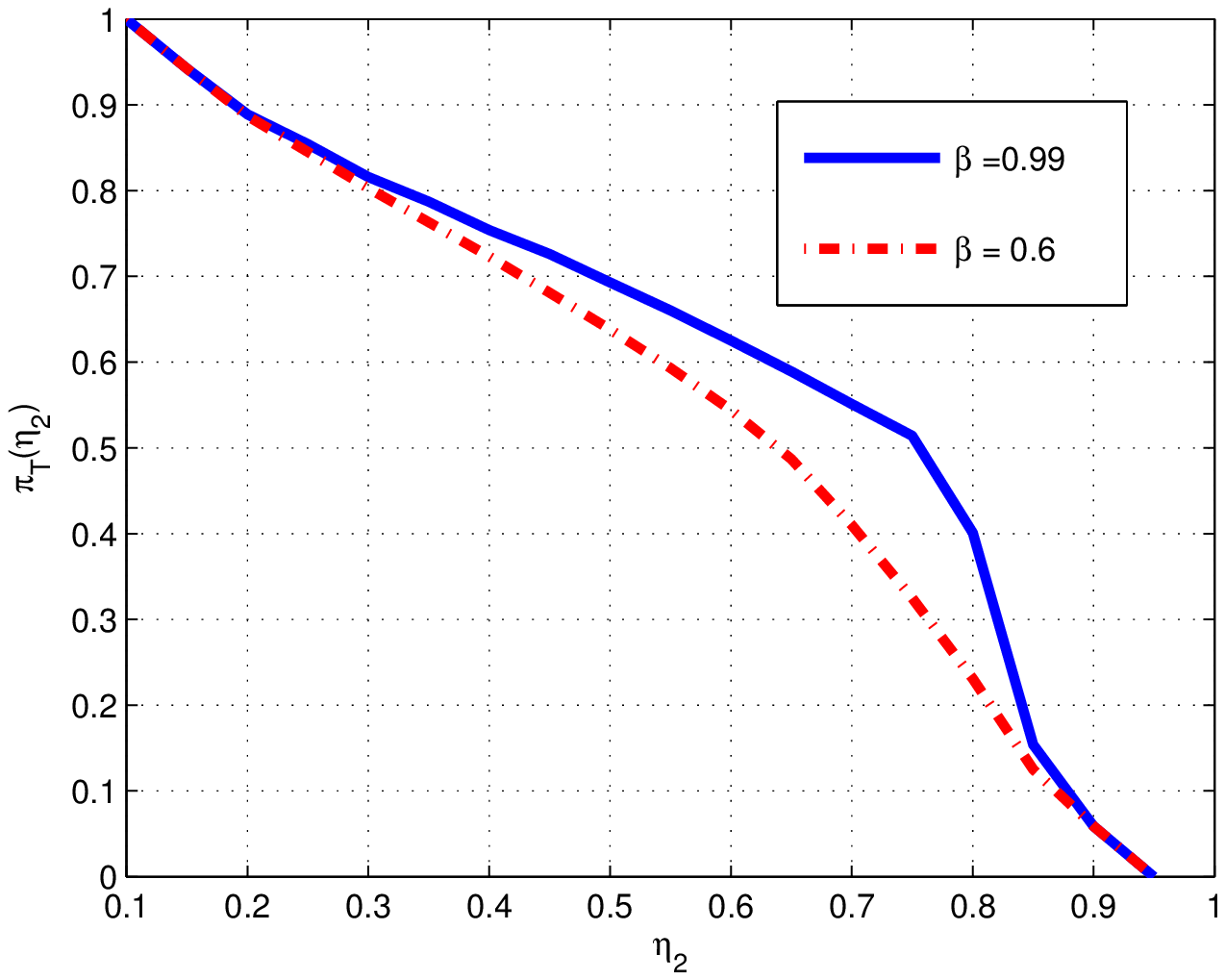} 
    \end{tabular}
    
  \end{center}
  \caption{$\pi_T(\eta_2)$ is plotted for $\beta = 0.6$ and $\beta=0.99.$
    The left plot is for the same set of same parameters as in 
    Fig.~\ref{plots:VS-VNS-set1} whereas the right plot uses 
    $\rho_0=\eta_0 =0.1,$ $\rho_1 =\eta_1 =0.95,$ $\mu_0=
  \lambda_0=0.9,$ and $\mu_1 = \lambda_1 =0.1.$
  }
  \label{plots:pi_T}
\end{figure*}

\end{remark}

\section{Explicit calculation of the Whittle index for the class 
of threshold policies}
\label{sec:threshold}

Recall Conjecture~\ref{conj:threshold} on a threshold policy for the
single-arm hidden Markov bandit. For the cases when the conjecture is
true, we can use the definition of the Whittle index for an arm and
explicitly evaluate it.  The calculations though are tedious and
require us to exercise care in enumerating the various cases. This is
because the properties of the $\gamma$s in Property~\ref{prop:gammas}
depend on the ordering of $\mu$s and $\lambda$s. In the following we
will consider, for the sake of an example, one case $\lambda_0=\mu_0 >
\mu_1 = \lambda_1.$ The other cases have similar calculations and will
be omitted here. We will also continue to assume that $0 < \rho_0 <
\rho_1 < 1.$

\begin{figure}[b]
\begin{center}
\begin{tikzpicture}[scale = 1.0]
\draw[->, thick] (0,0)  to (8,0);
\draw[red, thick] (0.3,-0.1)  to (0.3,0.1);
\draw[red, thick] (7.0,-0.1)  to (7.0,0.1);
\draw[blue, thick] (1.0,-0.1)  to (1.0,0.1);
\draw[blue, thick] (1.8,-0.1)  to (1.8,0.1);
\draw[blue, thick] (3.2,-0.1)  to (3.2,0.1);
\draw[blue, thick] (4.8,-0.1)  to (4.8,0.1);
\draw[blue, thick] (5.8,-0.1)  to (5.8,0.1);
\node (a) at (0.3,-0.3) {0};   
\node (d) at (7.0,-0.3) {$1$};
\node (e) at (8,-0.3) {$\pi$};
\node (f) at (1.0,-0.3) {$\mu_1$};
\node (g) at (1.8,-0.3) {$\gamma_{1,\infty}$};
\node (h) at (3.2,-0.3) {$\gamma_{2,\infty}$};
\node (i) at (4.8,-0.3) {$\gamma_{0,\infty}$};
\node (j) at (5.8,-0.3) {$\mu_0$};
\node (k) at (0.6,0.6) {$A_1$};
\node (l) at (1.9,0.6) {$A_2$};
\node (m) at (3.8,0.6) {$A_3$};
\node (n) at (5.2,0.6) {$A_4$};
\node (o) at (6.3,0.6) {$A_5$};
\draw[<-, red, thick] (0.3,0.3)  to (0.7,0.3);
\draw[->, red, thick] (0.7,0.3)  to (1.0,0.3);
\draw[<-, blue, thick] (1,0.3)  to (2.5,0.3);
\draw[->, blue, thick] (2.5,0.3)  to (3.2,0.3);
\draw[<-, magenta, thick] (3.2,0.3)  to (3.8,0.3);
\draw[->, magenta, thick] (3.8,0.3)  to (4.8,0.3);
\draw[<-, black, thick] (4.8,0.3)  to (5.2,0.3);
\draw[->, black, thick] (5.2,0.3)  to (5.8,0.3);
\draw[<-, red, thick] (5.8,0.3)  to (6.2,0.3);
\draw[->, red, thick] (6.2,0.3)  to (7.0,0.3);
\end{tikzpicture}
\end{center}

\caption{The different cases to calculate $W(\pi)$ in
  Section~\ref{sec:threshold}}
\label{fig:areas-for-WI}
\end{figure}
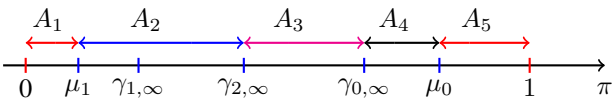

For $i=0, 1,2,$ define $\gamma_i^{(0)}(\pi) = \pi,$ $\gamma_i^{(k)} :=
\gamma_i\left( \gamma_i^{(k-1)}(\pi) \right), $ and $\gamma_{i,\infty}
:= \lim_{k \rightarrow \infty} \gamma_{i}^{(k)}(\pi).$ We can show
that $0 < \mu_1 < \gamma_{1,\infty} < \gamma_{2,\infty} <
\gamma_{0,\infty} < \mu_0 < 1.$ See Fig.~\ref{fig:areas-for-WI}. The
interval $(0,1),$ the range of $\pi$ is divided into five regions,
denoted $A_1,\ldots, A_5,$ as shown in Fig.~\ref{fig:areas-for-WI}.

\begin{enumerate}
\item For $\pi \in A_1,$  $W(\pi) = \rho(\pi).$
\item For $\pi \in A_2,$ we  will have the following cases
  \begin{enumerate}
  \item If $\gamma_1(\pi) \geq \pi,$ then $W(\pi) = \rho(\pi).$
  \item If $\pi > \gamma_1(\pi),$ $\pi \leq \gamma_0(\pi),$
    $\gamma_0(\gamma_1(\pi)) > \pi,$ and $\gamma_1^{(2)}(\pi) \geq  \pi$ then  
    \begin{eqnarray*}
      W(\pi) =  \frac{(1-\beta)\left[ \rho(\pi) +\beta \rho(\pi) 
          \rho(\gamma_1(\pi)) \right]}{\left(1-\beta 
        \left[1- \rho(\pi) (1-\beta) \right] \right)} .
    \end{eqnarray*}  
  \item If $\pi > \gamma_1(\pi),$ $\pi \leq \gamma_0(\pi),$
    $\gamma_0(\gamma_1(\pi)) > \pi,$ and $\gamma_1^{(2)}(\pi) < \pi,$ then  
    \begin{eqnarray*}
      W(\pi) = \frac{(1-\beta) C_1}{[1-(C_2 + C_3+ C_4)]},
    \end{eqnarray*}
    where 
    \begin{eqnarray*}
      C_1 &=& \sum_{l = 0}^{\tau_1-1} \beta^l \prod_{j=0}^l
      \rho\left(\gamma_1^{(j)}(\pi)\right) \\
      C_2 &=& \beta^{\tau_1} \prod_{j=0}^{\tau_1-1}
      \rho\left(\gamma_1^{(j)}(\pi)\right) \\
      C_3 &=& \sum_{l=1}^{\tau_1-1} \beta^{l+1} \prod_{j=1}^{l}
      \rho\left( \gamma_1^{(j-1)}(\pi)\right) \left(1-
      \rho\left(\gamma_1^{(l)}(\pi)\right)\right) \\ 
      C_4 &=& \beta( 1- \rho(\pi))\\ 
      \tau_1 &:=& \inf\left\{k \geq 1: \gamma_1^{(k)}(\pi) \geq \pi
      \right\}.
    \end{eqnarray*}
  \item If $\pi > \gamma_1(\pi),$ $\gamma_0(\pi) \geq \pi,$
    $\gamma_0(\gamma_1(\pi)) < \pi$ and $\gamma_1^{(2)}(\pi) < \pi$ 
    then $W(\pi)$ is obtained numerically by performing the value iteration 
    till convergence. 
  \end{enumerate}
\item For $\pi \in A_3$ then the Whittle index is obtained via 
  numerical computation as described above.
\item For $\pi \in A_4,$ $W(\pi) = \rho(\pi) + \beta \gamma_2(\pi)
  (m-1).$
\item For $\pi \in A _5 ,$ then
  \begin{eqnarray*}
    W(\pi) = m \pi \left(1 - \beta (\lambda_0 - \lambda_1) \right) +
    (1-\beta)c - \beta \lambda_1 m.
  \end{eqnarray*}
   where $m  =  \frac{\rho_0 - \rho_1}{ 1- \beta (\mu_0 -\mu_1)},$  
$c  =  \frac{\rho_1 + \frac{\beta \mu_1 (\rho_0 - \rho_1)}  {1 - \beta (\mu_0 - \mu_1)}}{1 - \beta}. $
\end{enumerate}

We now provide a brief description of the key steps in obtaining the
preceding expressions. The key idea is of course to solve
$V_{S}(\pi,\eta_2)=V_{NS}(\pi,\eta_2)$ for $\eta_2.$ This solution
is $W(\pi).$ In general, $V_{S}(\pi,\eta_2)$ and $V_{NS}(\pi, \eta_2)$ do
not have closed form expressions. The key step is to show that for
fixed $\eta_2,$ both $V_S(\pi, \eta_2)$ and $V_{NS}(\pi,\eta_2)$ have
at most three connected components for fixed $\eta_2.$ This fact, and
the properties of the $\gamma$s are then used to solve for $\eta_2.$
For example, for $\pi \in A_1,$ we have $0 \leq \pi \leq \mu_1,$
$\gamma_0(\pi), \gamma_1(\pi) \geq \pi$ and $V_S(\pi,\eta_2)=
\rho(\pi) + \beta \frac{\eta_2}{1-\beta}$ and $V_{NS}(\pi, \eta_2) =
\frac{\eta_2}{1-\beta}.$ Equating $V_S(\pi,\eta_2)$ and
$V_{NS}(\pi,\eta_2)$ at $\pi = \pi_T$ and solving for $\eta_2,$ we get
$\eta_2 = \rho(\pi) = W(\pi).$ The other closed form expressions are
similarly obtained. For the two cases for which we need to obtain
$W(\pi)$ numerically, such a simplification is not possible.

\section{Concluding Remarks}
\label{sec:discuss}

Several interesting prospects for future work are open. We would of
course like to know for sure if the single armed bandit indeed has a
single threshold sampling policy. As we mention in the appendix, the
complexity of the $\gamma_i$s makes such a proof hard and the `usual'
techniques that have been used in the literature do not appear to be
useful. The restriction on $\beta$ in the main results are in the same
spirit as that of \cite{White79}. The approximation is introduced
here.

Since we do not have a closed-form expression for $V(\pi)$ and
$W(\pi),$ provably good approximations may be sought. Also, since the
Whittle index based policy is itself suboptimal, we could seek other
suboptimal policies that can provide guarantees on the approximation
to optimality.

\bibliographystyle{IEEEbib}

\bibliography{restless-bandits}

\begin{thebibliography}{10}

\bibitem{Gilbert60}
E.~N. Gilbert,
\newblock ``Capacity of a {Burst-Noise} {Channel},''
\newblock {\em Bell {S}ystem {T}echnical {J}ournal}, vol. 39, no. 5, pp.
  1253--1265, 1960.

\bibitem{Papadimitriou99}
C.~H. Papadimitriou and J.~H. Tsitsiklis,
\newblock ``The complexity of optimal queueing network control,''
\newblock {\em Mathematics of Operations Research}, vol. 24, no. 2, pp.
  293--305, May 1999.

\bibitem{Nino-Mora08}
J.~Ni{\~{n}o}-Mora,
\newblock ``An index policy for dynamic fading-channel allocation to
  heterogeneous mobile users with partial observations,''
\newblock in {\em Proceedings of the Conference on Next Generation Internet
  Networks}, April 2008, pp. 231--238.

\bibitem{Nino-Mora09}
J.~Ni{\~{n}o}-Mora,
\newblock ``A restless bandit marginal productivity index for opportunistic
  spectrum access with sensing errors,''
\newblock in {\em Proceedings of Conference on Network Control Optimization
  {(NET-COOP), LNCS 5894}}, 2009, pp. 60--74.

\bibitem{LiuZhao10}
K.~Liu and Q.~Zhao,
\newblock ``Indexability of restless bandit problems and optimality of
  {W}hittle index for dynamic multichannel access,''
\newblock {\em IEEE Transactions Information Theory}, vol. 56, no. 11, pp.
  5557--5567, November 2010.

\bibitem{Lott10}
C.~Lott and D.~Teneketzis,
\newblock ``On the optimality of the index rule in multichannel allocation for
  single-hop mobile networks with multiple service classes,''
\newblock {\em Probability in the Engineering and Information Sciences}, vol.
  14, pp. 259--297, 2010.

\bibitem{Zhao08}
Q.~Zhao, B.~Krishnamachari, and K.~Liu,
\newblock ``On myopic sensing for multi-channel opportunistic access:
  structure, optimality, and performance,''
\newblock {\em IEEE Transactions on Wireless Communication}, vol. 7, no. 12,
  pp. 5431--5440, December 2008.

\bibitem{Wang13}
K.~Wang, L.~Chen, and Q.~Liu,
\newblock ``On optimality of myopic sensing policy with imperfect sensing in
  multi-channel opportunistic access,''
\newblock {\em IEEE Transactions on Communications}, vol. 61, no. 9, pp.
  3854--3862, September 2013.

\bibitem{Wang14}
K.~Wang, L.~Chen, and Q.~Liu,
\newblock ``On optimality of myopic policy for opportunistic access with
  nonidentical channels and imperfect sensing,''
\newblock {\em IEEE Transactions on Vehicular Technology}, vol. 63, no. 5, pp.
  2478--2483, June 2014.

\bibitem{Ouyang12}
W.~Ouyang, A.~Eyrilmaz, and N.~Shroff,
\newblock ``Asymptotically optimal downlink scheduling over {M}arkovian fading
  channels,''
\newblock in {\em Proceedings of {IEEE} {INFOCOM}}, 2012, pp. 1224--1232.

\bibitem{Zhao07}
Q.~Zhao, L.~Tong, A.~Swami, and Y.~Chen,
\newblock ``Decentralized cognitive {M}{A}{C} for opportunistic spectrum access
  in ad hoc networks: {A} {P}{O}{M}{D}{P} framework,''
\newblock {\em IEEE Journal on Selected Areas in Communications}, vol. 25, no.
  3, pp. 589--600, April 2007.

\bibitem{Chen08}
Y.~Chen, Q.~Zhao, and A.~Swami,
\newblock ``Joint design and separation principle for opportunistic spectrum
  access in the presence of sensing errors,''
\newblock {\em IEEE Transactions on Information Theory}, vol. 54, no. 5, pp.
  2053--2071, May 2008.

\bibitem{Li13}
C.~Li and M.~J. Neely,
\newblock ``Network utility maximization over partially observable {M}arkovian
  channels,''
\newblock {\em Performance Evaluation}, vol. 70, no. 7--8, pp. 528--548, July
  2013.

\bibitem{Whittle88}
P.~Whittle,
\newblock ``Restless bandits: {A}ctivity allocation in a changing world,''
\newblock {\em Journal of Applied Probability}, vol. 25, no. A, pp. 287--298,
  1988.

\bibitem{Veatch96}
M.~H. Veatch and L.~M. Wein,
\newblock ``Scheduling a make-to-stock queue: {I}ndex policies and hedging
  points,''
\newblock {\em Operations Research}, vol. 44, no. 4, pp. 634--647, July-August
  1996.

\bibitem{Ny08}
J.~L. Ny, M.~Dahleh, and E.~Feron,
\newblock ``Multi-{UAV} dynamic routing with partial observations using
  restless bandit allocation indices,''
\newblock in {\em Proceedings of American Control Conference (ACC)}, 2008, pp.
  4220--4225.

\bibitem{Ouyang11}
W.~Ouyang, S.~Murugesan, A.~Eyrilmaz, and N.~Shroff,
\newblock ``Exploiting channel memory for joint estimation and scheduling in
  downlink networks,''
\newblock in {\em Proceedings of {IEEE INFOCOM}}, 2011.

\bibitem{Avrachenkov13}
K.~Avrachenkov, U.~Ayesta, J.~Doncel, and P.~Jacko,
\newblock ``Congestion control of {TCP} flows in {I}nternet routers by means of
  index policy,''
\newblock {\em Computer Networks}, vol. 57, no. 17, pp. 3463--3478, 2013.

\bibitem{Avrachenkov16}
K.~Avrachenkov and V.~S. Borkar,
\newblock ``Whittle index policy for crawling ephemeral content,''
\newblock {\em {IEEE} Transactions on Control of Network Systems}, 2016,
  DOI:{10.1109/TCNS.2016.2619066}.

\bibitem{Nino-Mora01}
J.~E. Ni{\~{n}o}-Mora,
\newblock ``Restless bandits, partial conservation laws and indexability,''
\newblock {\em Advances in Applied Probability}, vol. 33, pp. 76--98, 2001.

\bibitem{Ross71}
S.~M. Ross,
\newblock ``Quality control under {M}arkovian deterioration,''
\newblock {\em Management Science}, vol. 17, no. 9, pp. 587--596, May 1971.

\bibitem{Sernik91a}
E.~L. Sernik and S.~I. Marcus,
\newblock ``On the computation of optimal cost function for discrete time
  {M}arkov models with partial observations,''
\newblock {\em Annals of Operations Research}, vol. 29, pp. 471--512, 1991.

\bibitem{Sernik91b}
E.~L. Sernik and S.~I. Marcus,
\newblock ``Optimal cost and policy for a {M}arkovian replacement problem,''
\newblock {\em Journal of Optimization Theory and Applications}, vol. 71, no.
  1, pp. 403--406, October 1991.

\bibitem{Hughes80}
J.~S. Hughes,
\newblock ``A note on quality control under {M}arkovian deterioration,''
\newblock {\em Operations Research}, vol. 28, no. 2, pp. 421--424, March-April
  1980.

\bibitem{Krishnamurthy01}
V.~Krishnamurthy and R.~J. Evans,
\newblock ``Hidden {M}arkov model for multiarm bandits: a methodology for beam
  scheduling in multitarget tracking,''
\newblock {\em IEEE Transactions on Signal Processing}, vol. 49, no. 12, pp.
  2893--2908, December 2001.

\bibitem{Astrom69}
K.~J. Astrom,
\newblock ``Optimal control of {M}arkov processes with incomplete state
  information {II.} {T}he convexity of loss function,''
\newblock {\em Mathematical Analysis and Applications}, vol. 26, no. 2, pp.
  403--406, May 1969.

\bibitem{BertsekasV195}
D.~P. Bertsekas,
\newblock {\em Dynamic Programming and Optimal Control}, vol.~1,
\newblock Athena Scientific, Belmont, Massachusetts, 1st edition, 1995.

\bibitem{BertsekasV295}
D.~P. Bertsekas,
\newblock {\em Dynamic Programming and Optimal Control}, vol. 2nd,
\newblock Athena Scientific, Belmont, Massachusetts, 1st edition, 1995.

\bibitem{White79}
C.~C.~White III,
\newblock ``Optimal control-limit strategies for a partially observed
  replacement problem,''
\newblock {\em International Journal of System Science}, vol. 10, no. 3, pp.
  321--331, 1979.

\bibitem{Clarke90}
F.~H. Clarke,
\newblock {\em Optimization and Nonsmooth Analysis},
\newblock SIAM, Philadelphia, 1990.

\end{thebibliography}
 
\newpage

\appendix

\subsection{Proof of Lemma~ \ref{lemma:background}}
\label{app:background-lemma}

Let $x,y \in \Re_+^n$ and $0 \leq \alpha \leq 1.$ Then we have the
following.

\begin{eqnarray*}
  && g\left(\alpha x + (1-\alpha)x\right)\\
   & & = \  ||\alpha x + (1-\alpha)
  y||_1 f\left(\frac{\alpha x + (1-\alpha)y}{||\alpha x +
    (1-\alpha)y||_1} \right) \\
  && =  \ ||\alpha x + (1-\alpha) y||_1 f\left(\frac{\alpha ||x||_1}
  {||\alpha x + (1-\alpha)y||_1} \frac{x}{||x||_1} \right. \\
 && \ \ \ \left. + \frac{(1-\alpha) ||y||_1}{||\alpha x +
    (1-\alpha)y||_1}   \frac{y}{||y||_1} \right) \\
  & & \leq \ ||\alpha x + (1-\alpha) y||_1 \left[ 
    \frac{\alpha ||x||_1} {||\alpha x + (1-\alpha)y||_1} 
    f\left(\frac{x}{||x||_1} \right) \right. \\
   && \ \ \ \left. + \frac{(1-\alpha) ||y||_1}{||\alpha x +
      (1-\alpha)y||_1}  f\left(  \frac{y}{||y||_1} \right) \right]\\
    && = \  \alpha ||x||_1  f\left(\frac{x}{||x||_1} \right)
    + (1-\alpha) ||y||_1 f\left(  \frac{y}{||y||_1} \right) \\
    && = \ \alpha g(x) + (1-\alpha)g(y)
  \end{eqnarray*}
The inequality in the fifth line follows from convexity of $f.$ 

\subsection{Proof of Lemma~\ref{lemma:Vs-of-pi-rho2}}
\label{app:lemma:Vs-of-pi-rho2}

For part~(1), we first prove that $V(\pi)$ is convex by induction and
use this to show that $V_S(\pi)$ and $V_{NS}(\pi)$ are also
convex. Let
\begin{eqnarray}
  V_1(\pi) & = & \max \left\{\pi \rho_0 + (1-\pi) \rho_1, \eta_2 \right \} , 
  \nonumber \\
  V_{n+1}(\pi) & = & \max\left \{ \eta_2+ \beta V_n(\gamma_2(\pi)),
  \right. \rho(\pi) + \nonumber \\
  && \hspace{-30pt} \left. \beta \left[ \rho(\pi) V_n(\gamma_1(\pi)) + 
      (1-\rho(\pi)) V_n(\gamma_0(\pi)) \right] \right\} . 
  \label{eqn:V-n+1-of-pi}
\end{eqnarray}
Now define 
\begin{eqnarray}
  b_0 & := & \left[\pi \mu_0 (1-\rho_0)+(1-\pi)\mu_1(1-\rho_1), 
  \right. \nonumber\\
  && \ \left.  (1-\mu_0)(1-\rho_0)\pi + (1-\pi((1-\mu_1)(1-\rho_1) 
  \right]^T , \nonumber \\
  b_1 & := & \left[\pi \mu_0 \rho_0+ (1-\pi)\mu_1 \rho_1, \right. 
  \nonumber\\
  && \left. (1-\mu_0)\rho_0\pi + (1-\pi((1-\mu_1)\rho_1 \right]^T,
  \label{eqn:a-notation}
\end{eqnarray}
and  write \eqref{eqn:V-n+1-of-pi} as
\footnotesize{
\begin{eqnarray*}
  V_{S,n+1}(\pi) \hspace{-3pt} & = & \hspace{-3pt} ||b_1||_1 + \beta ||b_1||_1 
  V_n\left(\frac{b_1}{||b_1||_1}\right) 
  + \beta ||b_0||_1 
  V_n\left(\frac{b_0}{||b_0||_1}\right).
\end{eqnarray*}
}

\normalsize
\noindent
Here superscript $T$ denotes the transpose. Clearly, $V_1(\pi)$ is
linear and hence convex.  Making the induction hypothesis that
$V_n(\pi)$ is convex in $\pi,$ $V_{S,n+1}(\pi)$ and $V_{n+1}(\pi)$ are
convex from Lemma~\ref{lemma:background} and by induction $V_n(\pi)$
is convex for all $n.$ From Chapter~7 of \cite{BertsekasV195} and
Proposition 2.1 of Chapter~2 of \cite{BertsekasV295}, $V_n(\pi) \to
V(\pi)$ and hence $V(\pi)$ is convex, Further,
\begin{eqnarray*}
  V^{\prime \prime}_{NS}(\pi) & = & \beta V^{\prime \prime
  }(\gamma_2(\pi)) \left( \gamma_2^{\prime}(\pi) \right)^2 
\end{eqnarray*}
and hence $V_{NS}$ is also convex. Using the notation from
\eqref{eqn:a-notation}, we can write
\footnotesize{
\begin{eqnarray*}
  V_S(\pi) = ||b_1||_1 + \beta ||b_1||_1 
  V\left(\frac{b_1}{||a_1||b_1}\right)
  + \beta ||b_0||_1 
  V\left(\frac{b_0}{||b_0||_1}\right) .
\end{eqnarray*}
}

\normalsize 

\noindent
The first term in the RHS above is clearly convex in $\pi.$ Since
$V(\pi)$ is convex, from Lemma~\ref{lemma:background}, the second and
third terms are also convex. Thus $V_S(\pi)$ is convex.

To prove the second part of the lemma we rewrite the recursion of
\eqref{eqn:V-n+1-of-pi} as follows.
\begin{eqnarray}
  V_{1}(\pi,\eta_2) & = & \max \{\rho(\pi), \eta_2\} \nonumber \\
  V_{n+1}(\pi,\eta_2) & =& \max \left\{\eta_2 + 
    \beta V_{n}(\gamma_2(\pi),\eta_2), 
    \rho(\pi) + \right. \nonumber \\
  && \hspace{-70pt} \left.  \rho(\pi) \beta V_{n}(\gamma_1(\pi),\eta_2) + 
    \left(1-\rho(\pi)\right) \beta V_{n}(\gamma_0(\pi),\eta_2) 
  \right\}
  \label{algo:iterAlgo}
\end{eqnarray}
Here we have made explicit the dependence of $V(\pi)$ on $\eta_2.$ We
see that $V_{1}(\pi,\eta_2)$ is monotone non decreasing and convex in
$\eta_2.$ Make the induction hypothesis that for a fixed $\pi,$
$V_{n}(\pi,\eta_2)$ is monotone non decreasing and convex in $\eta_2.$
Then, in \eqref{algo:iterAlgo}, the first term of the $\max$ function
is the sum of two non decreasing convex functions of $\eta_2.$ The
second term is a constant plus a convex sum of two non decreasing
convex functions of $\eta_2.$ Thus it is also non decreasing and
convex in $\eta_2.$ The max operation preserves convexity. Thus
$V_{n+1}(\pi, \eta_2)$ is also non decreasing and convex in $\eta_2$
and by induction, all $V_n(\pi,\eta_2)$ are non decreasing and convex
in $\eta_2.$ As in the first part of the lemma, $V_n(\pi, \eta_2) \to
V(\pi, \eta_2)$ and this completes the proof for $V(\pi).$ From
\eqref{eqn:VS,VNS,V}, the assertion on $V_S(\pi)$ and $V_{NS}(\pi)$
follows.

\subsection{ Proof that $V_{\beta}(\pi,\eta_2)$ is increasing in $\beta$}
\label{subsec:proof-V-increase-beta}

If $\beta_a > \beta_b,$ we need to show that $V_{\beta_a}(\pi, \eta_2) >
V_{\beta_b}(\pi, \eta_2).$ Like in earlier proofs, we use an induction 
argument. Let 
\begin{eqnarray*}
  V_{S,\beta,1}(\pi,\eta_2) &:=& \rho(\pi), \\
  V_{NS,\beta,1}(\pi,\eta_2) &:=& \eta_2, \\
  V_{\beta,1}(\pi,\eta_2) &:=& \max 
  \{ V_{S,\beta,1}(\pi,\eta_2) , V_{NS,\beta,1}(\pi,\eta_2) \}, 
\end{eqnarray*}
and define
\begin{eqnarray}
  V_{S,\beta,n+1}(\pi,\eta_2) &:=& \rho(\pi) + \beta \left[ \rho(\pi) 
    V_{\beta,n}(\gamma_1(\pi),\eta_2) +  \right. \nonumber \\
    && \hspace{10pt} \left. (1 -\rho(\pi)) 
    V_{\beta,n}(\gamma_0(\pi),\eta_2) \right] , \nonumber \\
  V_{NS,\beta,n+1}(\pi,\eta_2) &:=& \eta_2 + \beta 
  V_{\beta,n}(\gamma_2(\pi),\eta_2)  , \nonumber \\
  V_{\beta,n+1}(\pi,\eta_2) &:=& \max \{ V_{S,\beta,n}(\pi,\eta_2) , 
  V_{NS,\beta,n}(\pi,\eta_2) \}. 
  \nonumber \\
  \label{eqn:induction-for-Vbeta}
\end{eqnarray}

Clearly, $V_{S,\beta,1}(\pi,\eta_2),$ $V_{NS,\beta,1}(\pi,\eta_2)$ and
$V_{\beta,1}(\pi,\eta_2)$ are all increasing in $\beta.$

Now make the induction hypothesis that $V_{S,\beta,n}(\pi,\eta_2),$
$V_{NS,\beta,n}(\pi,\eta_2)$ and by inspection of
\eqref{eqn:induction-for-Vbeta} we see that
$V_{S,\beta,n+1}(\pi,\eta_2),$ $V_{NS,\beta,n+1}(\pi,\eta_2),$ and
$V_{\beta,n+1}(\pi,\eta_2)$ are all increasing in $\beta.$ Further,
like in the proof Lemma~\ref{lemma:Vs-of-pi-rho2}, we know that
\begin{eqnarray*}
  V_{n,S,\beta}(\pi,r_2) & \rightarrow & V_{S,\beta}(\pi,r_2), \\
  V_{n,NS,\beta}(\pi,r_2) & \rightarrow & V_{NS,\beta}(\pi,r_2), \\
  V_{n,\beta}(\pi,r_2) & \rightarrow & V_{\beta}(\pi,r_2),
\end{eqnarray*}
and the claim follows. 

\subsection{Proof of Lemma \ref{lemma:diff-value-S-NS-monotone}}

For a function $f:\Re \rightarrow \Re$ that is continuous and
Lipschitz, we specialize the notion of a generalized gradient (see,
e.g., \cite{Clarke90}) and define
\begin{displaymath}
  \partial f(x) := \mathsf{co}\{\partial f_L(x), \partial f_R(x) \}.
\end{displaymath}
Here $\partial f_L(x)$ and $\partial f_R(x)$ are, respectively, the
left and right derivatives of $f$ at $x$ and $\mathsf{co}\{ \}$
represents the convex hull. Many operations and properties of the
gradient follow through to the generalized gradient. In particular,
the following will be used.
\begin{itemize}
\item (Chain rule) If $f(x) = (g \circ h) (x),$ with $g: \Re
  \rightarrow \Re$ and $h: \Re \rightarrow \Re,$ $h$ is differentiable
  and $g$ is convex, then
  \begin{eqnarray*}
    \partial f(x) = \partial g(h(x)) \frac{d h(x)}{dx}. 
  \end{eqnarray*}
\item (Mean value theorem) If $x, y \in \Re,$ $f$ is Lipschitz on an
  open set containing line segment $[x,y],$ then there exists a point
  $u \in (x,y)$ such that
  \begin{eqnarray*}
    f(y) - f(x) \in (y-x)\cdot \partial f(u).
  \end{eqnarray*} 
\end{itemize}

First, for any $0 \leq \pi_1 \leq \pi_2 \leq 1,$ we obtain a bound on
$|V(\pi_2) -V(\pi_1)|.$ The proof will follow the iterative technique
as in Appendix~\ref{subsec:proof-V-increase-beta}.  Define $\kappa_1
:= (1-\beta \vert \mu_0 - \mu_1 \vert)^{-1}.$
%

%\section{Lemma $5$}
\begin{lemma}
  For a fixed $\eta_2,$ $\beta_2 \in (0,1],$ and $0 \leq \pi_1 \leq
    \pi_2 \leq 1,$ if either $0< \mu_0-\mu_1 \leq \frac{1}{2}$ or $0<
    \mu_1-\mu_0 <1$ is true, then
    \begin{eqnarray*}
      \big\vert V(\pi_2) - V(\pi_1) \big \vert \leq \kappa_1 \vert \rho_1-
      \rho_0\vert \vert\pi_2-\pi_1\vert.
    \end{eqnarray*}      
    \label{lemma:diff-value-fun-bdd}
\end{lemma}

\begin{IEEEproof}
  We present the calculations for $0<\mu_0-\mu_1 \leq \frac{1}{2}.$
  The calculations for $0< \mu_1-\mu_0 <1$ are identical.
 
  \begin{enumerate}
  \item Let $V_1(\pi) = \max \{\rho(\pi), \eta_2 \},$ recall that
    $\rho(\pi) = \pi (\rho_0 -\rho_1) + \rho_1$ and $\rho_0 < \rho_1.$
    The generalized gradient of $V_1$ at $\pi \in [0,1]$ is
    \begin{eqnarray*}
      \partial V_1(\pi) & = & \mathsf{co} \{\rho_0 - \rho_1, 0 \} =
               [\rho_0-\rho_1, 0] \\ 
               & \subset & [-\kappa_1(\rho_1 -\rho_0), \kappa_1(\rho_1
                 -\rho_0)].
    \end{eqnarray*}
    
     \item Applying the the chain rule on the generalized gradient, we
       get
    {\small{
        \begin{eqnarray}
          \partial V_{S,n+1}(\pi) & = & \mathrm{co}\bigg\{(\rho_0 -
          \rho_1) \nonumber \\
          && \hspace{-35pt}
          + \beta (\rho_1 - \rho_0)\left[V_{n}(\gamma_0(\pi)) -
            V_{n}(\gamma_1(\pi)) \right]  \nonumber \\
          && \hspace{-35pt}
          + \beta  \frac{\rho_1 \rho_0 (\mu_0 -     
            \mu_1)}{(\rho(\pi))} \partial V_{n}((\gamma_1(\pi)))  \nonumber\\
          && \hspace{-35pt}
          + \beta \frac{(1-\rho_1)(1-\rho_0)(\mu_0 -
          \mu_1)}{(1-\rho(\pi))} \partial V_{n}(\gamma_0(\pi)) \bigg\}
          \nonumber \\
          \label{eq:partial-V-S-pi-nplusa}
        \end{eqnarray}
    }}

  \item We make the induction hypothesis that $\big\vert V_n(\pi_2) -
    V_n(\pi_1) \big \vert \leq \kappa_1 \vert \rho_1- \rho_0\vert
    \vert\pi_2-\pi_1\vert$ for all $0 \leq \pi_1 \leq \pi_2 \leq 1$
    and provide upper and lower bounds for $\partial V_{S,n+1}(\pi).$
  \item First, consider the upper bound. For $\mu_0 > \mu_1,$ from
    Property \ref{prop:gammas}, we see that for $0 \leq \pi \leq 1,$
    $\mu_0>\gamma_0(\pi) > \gamma_1(\pi) > \mu_1.$ Hence
    $\left(\gamma_0(\pi) - \gamma_1(\pi)\right) \leq (\mu_0 - \mu_1).$
    Using this and the mean value theorem for the generalized
    gradient, we obtain the following bound.
    \begin{eqnarray*}
      \big \vert 
      V_{n}(\gamma_0(\pi)) - V_{n}(\gamma_1(\pi)) \big \vert
      & \leq & \kappa_1(\rho_1 - \rho_0) (\mu_0 - \mu_1),
    \end{eqnarray*}
    Further, from the induction hypothesis, 
    \begin{eqnarray*}
      \partial V_{n}(\gamma_1(\pi)), \partial V_{n}(\gamma_0(\pi)) =  
               [-\kappa_1(\rho_1-\rho_0),\kappa_1(\rho_1-\rho_0)] 
    \end{eqnarray*}    
    Hence, using the observation that $\rho_0 \leq \rho(\pi) \leq
    \rho_1,$ and $(1-\rho_1) \leq (1-\rho(\pi)) \leq (1-\rho_0),$ with
    some calculations we can show that $\partial V_{S, n+1}(\pi)$ is
    upper bounded by
    \begin{eqnarray*}
      && (\rho_0 - \rho_1) + \beta \kappa_1(\rho_1 - \rho_0)^2 (\mu_0
      - \mu_1) \\
      && + \beta \rho_1 (\mu_0 - \mu_1) \kappa_1 (\rho_1 -
      \rho_0) \nonumber \\ 
      &&+ \beta (1-\rho_0)(\mu_0 - \mu_1)
      \kappa_1 (\rho_1 - \rho_0)
    \end{eqnarray*}
    which, after rearranging the terms becomes 
    \begin{eqnarray}
      && (\rho_1 - \rho_0) \kappa_1 \left( -1 + 4 \beta (\mu_0 -
      \mu_1) \right).
      \label{eq:partial-V-S-pi-UBa}
    \end{eqnarray}

    Now, since $0 < \mu_0 - \mu_1 \leq \frac{1}{2},$ we have $(-1 + 4
    \beta (\mu_0 - \mu_1)) \leq (-1 + 2 \beta ) \leq 1,$ and the upper
    bound becomes $\kappa_1(\rho_1 - \rho_0)$
   \item To obtain the lower bound, we substitute $\rho(\pi) \leq \rho_1$
    and $(1-\rho(\pi)) \leq (1 -\rho_0)$ in Eq.~
    \eqref{eq:partial-V-S-pi-nplusa}.  Using the induction hypothesis
    on $V_n(\pi),$ we can show that the lower bound is $- \kappa_1
    (\rho_1 - \rho_0).$

  \item From the preceding two steps we have
    \begin{eqnarray*}
      \partial V_{S,n+1}(\pi) \subseteq [
        -\kappa_1(\rho_1-\rho_0),\kappa_1(\rho_1-\rho_0)].
    \end{eqnarray*}
  \item Now consider generalized gradient of $V_{NS,n+1}(\pi)$
    w.r.t. $\pi.$ From equation \eqref{eqn:induction-for-Vbeta}, using
    properties of $\gamma_2(\pi)$ and the induction hypothesis on
    $V_n(\pi)$ with some algebra, we can obtain following inequality.
    \begin{eqnarray*}
      \partial V_{NS,n+1}(\pi) \subseteq [
        -\kappa_1(\rho_1-\rho_0),\kappa_1(\rho_1-\rho_0)].
    \end{eqnarray*}
  \item From the preceding two steps, we have %
    \begin{eqnarray*}
      \partial V_{n+1}(\pi) \subseteq [
        -\kappa_1(\rho_1-\rho_0),\kappa_1(\rho_1-\rho_0)].
    \end{eqnarray*}
    Thus, $\partial V_{n}(\pi) \subseteq [- \kappa_1 (\rho_1 -
      \rho_0), \kappa_1 (\rho_1 - \rho_0)]$ holds for every $n \geq 1$
    and $\pi \in [0,1].$ Also, $\lim_{n \rightarrow \infty }
    V_n(\pi) = V(\pi)$ converges uniformly.  Hence 
    \begin{eqnarray*}
      \partial V(\pi) = [-\kappa_1(\rho_1-\rho_0),\kappa_1
        (\rho_1-\rho_0)].
    \end{eqnarray*}
    Our claim follows.
    
  \end{enumerate}    
\end{IEEEproof}

We are now ready to prove 
Lemma~\ref{lemma:diff-value-S-NS-monotone}.  We consider the two cases
separately.

\textit{Case 1: $0<\mu_0 - \mu_1 \leq \frac{1}{5}$ and $\vert
  \lambda_0 - \lambda_1 \vert \leq \frac{1}{5}.$}
Define
\begin{eqnarray}
  d(\pi) := V_S(\pi) - V_{NS}(\pi).
  \label{eq:diff-V-S-V-NS}
\end{eqnarray}
The result is proved by showing that $\partial d(\pi) < 0.$ Consider
\begin{eqnarray*}
  \partial d(\pi) & = & \partial V_{S}(\pi) - \partial
  V_{NS}(\pi).
  \label{eq:partial-diff-V-S-V-NS}
\end{eqnarray*}
From the chain rule of the generalized gradient, we obtain 
\begin{eqnarray*}
  \partial V_{NS}(\pi) = \mathrm{co} \bigg\{ \beta \partial
  V(\gamma_2(\pi)) \gamma_2^{\prime}(\pi) \bigg\}.
\end{eqnarray*}
Further, we can show that $\partial V_{NS}(\pi)$ is lower bounded by
$- \beta \kappa_1 (\rho_1 - \rho_0) \vert \lambda_0 - \lambda_1 \vert$
and $\partial V_S(\pi)$ can be upper bounded by
\begin{eqnarray*}
(\rho_1 - \rho_0) \kappa_1 (1- \beta (\mu_0 - \mu_1)).
\end{eqnarray*}
Thus we can  upper bound  $\partial d(\pi)$ by 
\begin{eqnarray*}
 (\rho_1 - \rho_0) \kappa_1
  \left(- 1 + 4 \beta (\mu_0 - \mu_1) + \beta \vert \lambda_0 -
  \lambda_1 \vert \right).
\end{eqnarray*}
By our assumptions on $(\mu_0 - \mu_1)$ and $\vert\lambda_0 -
\lambda_1\vert,$ The upper bound on $\partial d(\pi)$ is less than
$0.$ Hence our claim follows.

\textit{Case 2: $0<\mu_1 - \mu_0 \leq \frac{1}{3}$ and $\vert
  \lambda_0 - \lambda_1 \vert \leq \frac{1}{3}.$}
Here, we can obtain following upper bound on $\partial d(\pi)$ using
similar tricks.
\begin{eqnarray*}
   (\rho_1 - \rho_0) \kappa_2 \left\{- 1 + 2 \beta (\mu_1 -
  \mu_0) + \beta \vert \lambda_0 - \lambda_1 \vert \right\}.
\end{eqnarray*}
From our assumptions on $(\mu_1 - \mu_0)$ and $\vert \lambda_0 -
\lambda_1\vert,$ we can show the upper bound on $\partial d(\pi)$ is
less than $0.$

This completes the proof.

\subsection{Sample Numerical Results for $V_S(\pi)$ and $V_{NS}(\pi)$ }
\label{app:numericals-vs-vns}
\begin{figure*}

  \begin{center}
    \begin{tabular}{cc}
      \includegraphics[width=0.8\columnwidth]{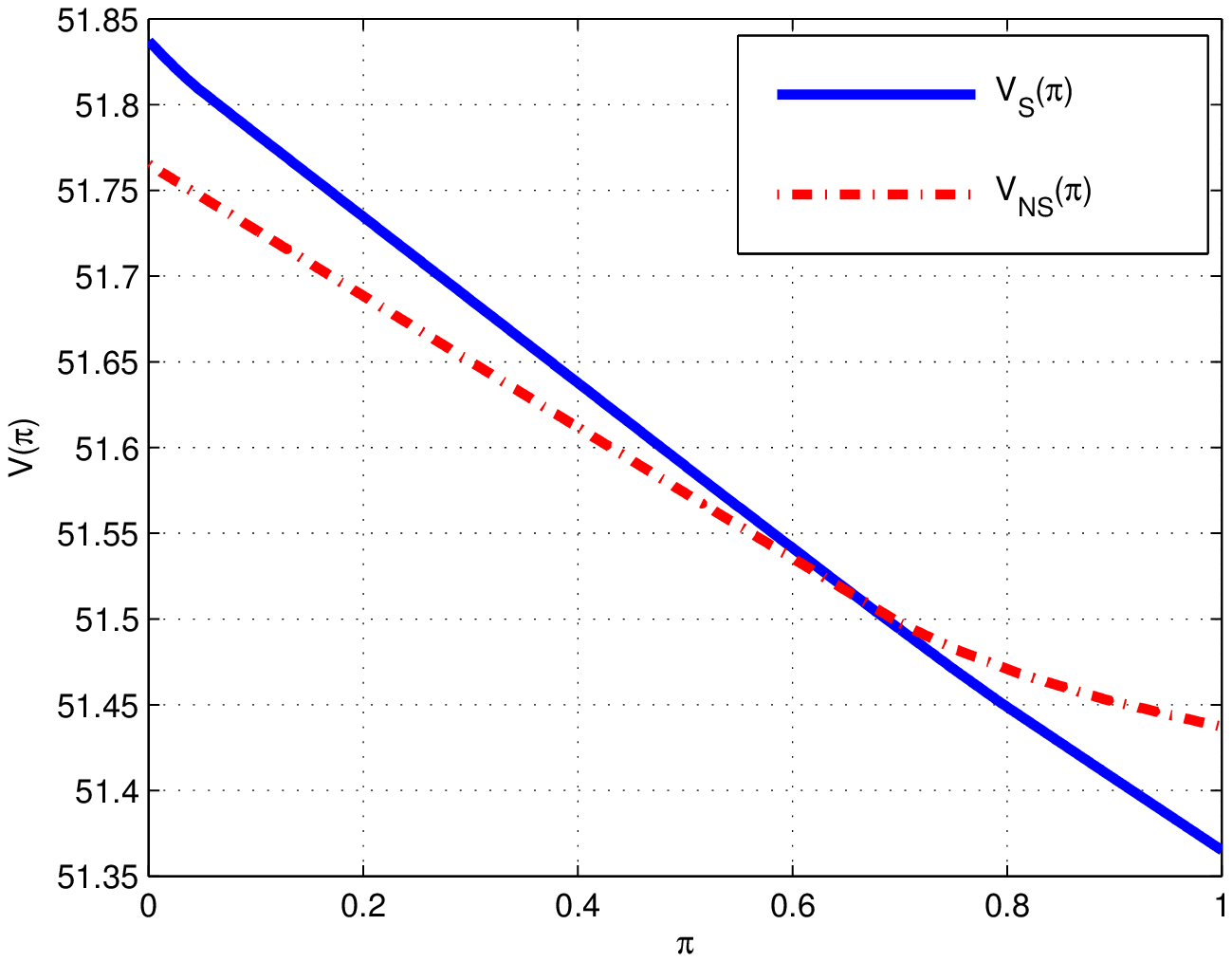}
      & 
      \includegraphics[width=0.8\columnwidth]{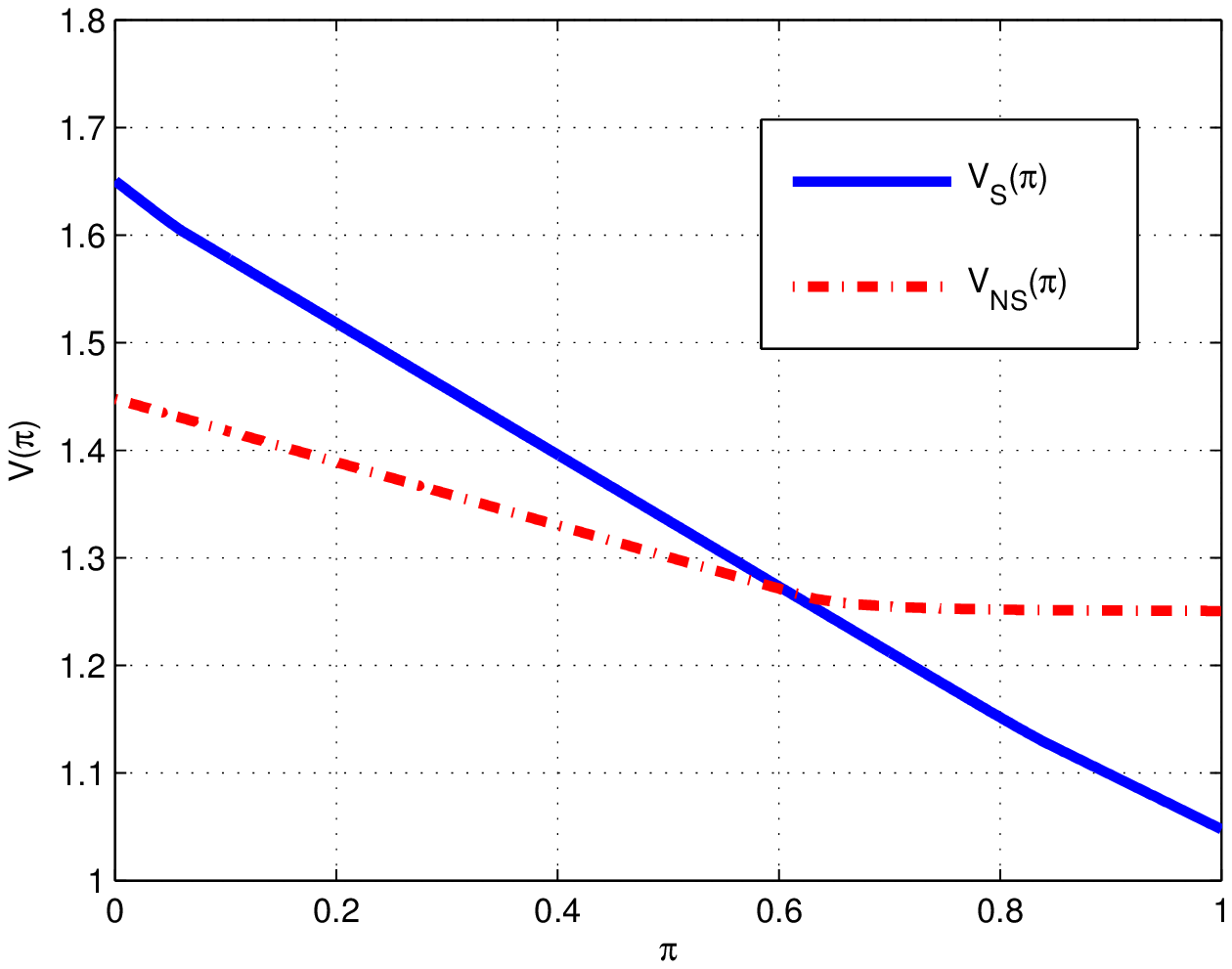} \\
      $\beta = 0.99;$ $\pi_T=0.673.$ &  $ \beta = 0.6; \pi_T=0.604.$ 
    \end{tabular}

    $\eta_2=0.5;$ $\pi^{\circ} = 0.5.$
  \end{center}

  \begin{center}
    \begin{tabular}{cc}
      \includegraphics[width=0.8\columnwidth]{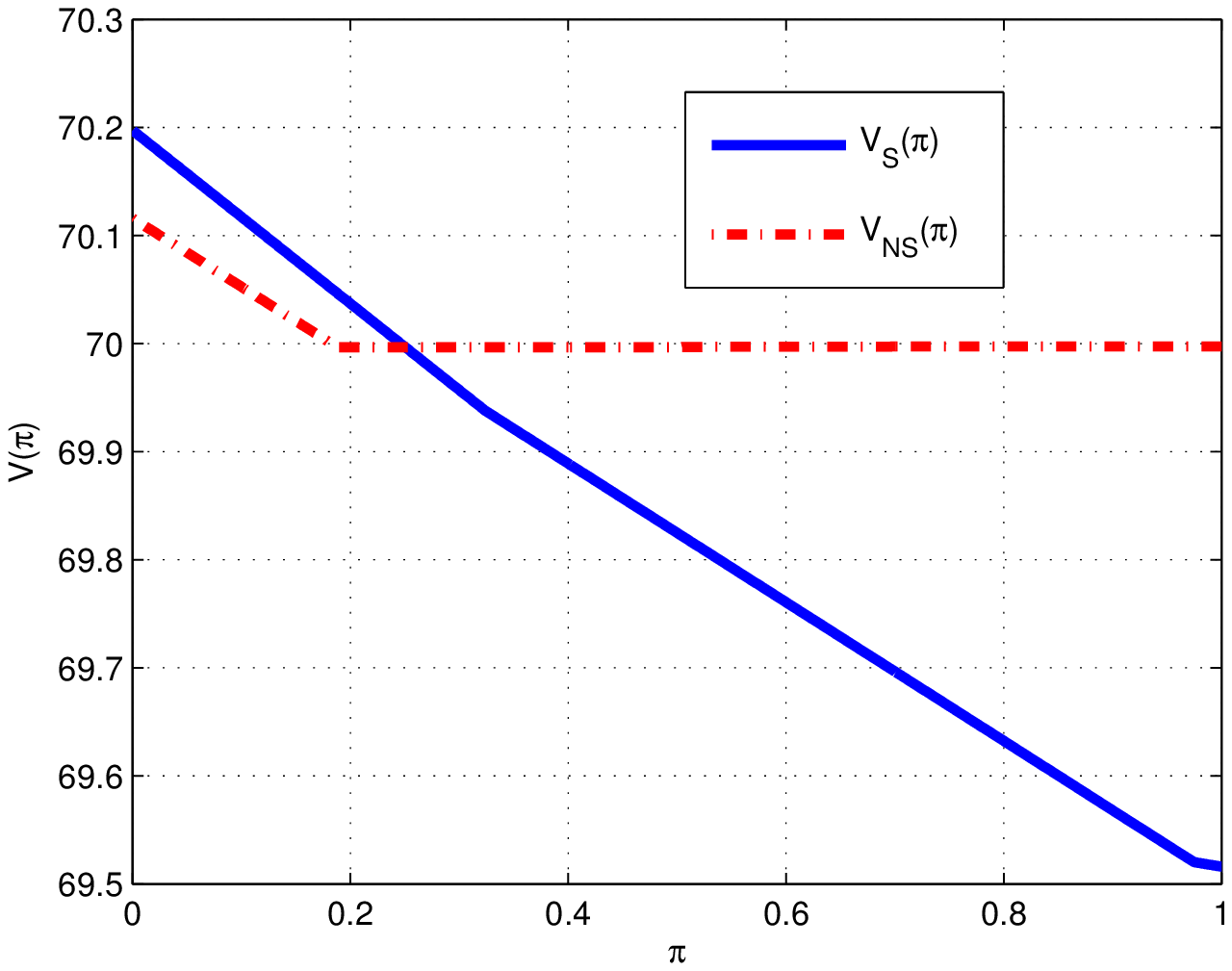}
      & 
      \includegraphics[width=0.8\columnwidth]{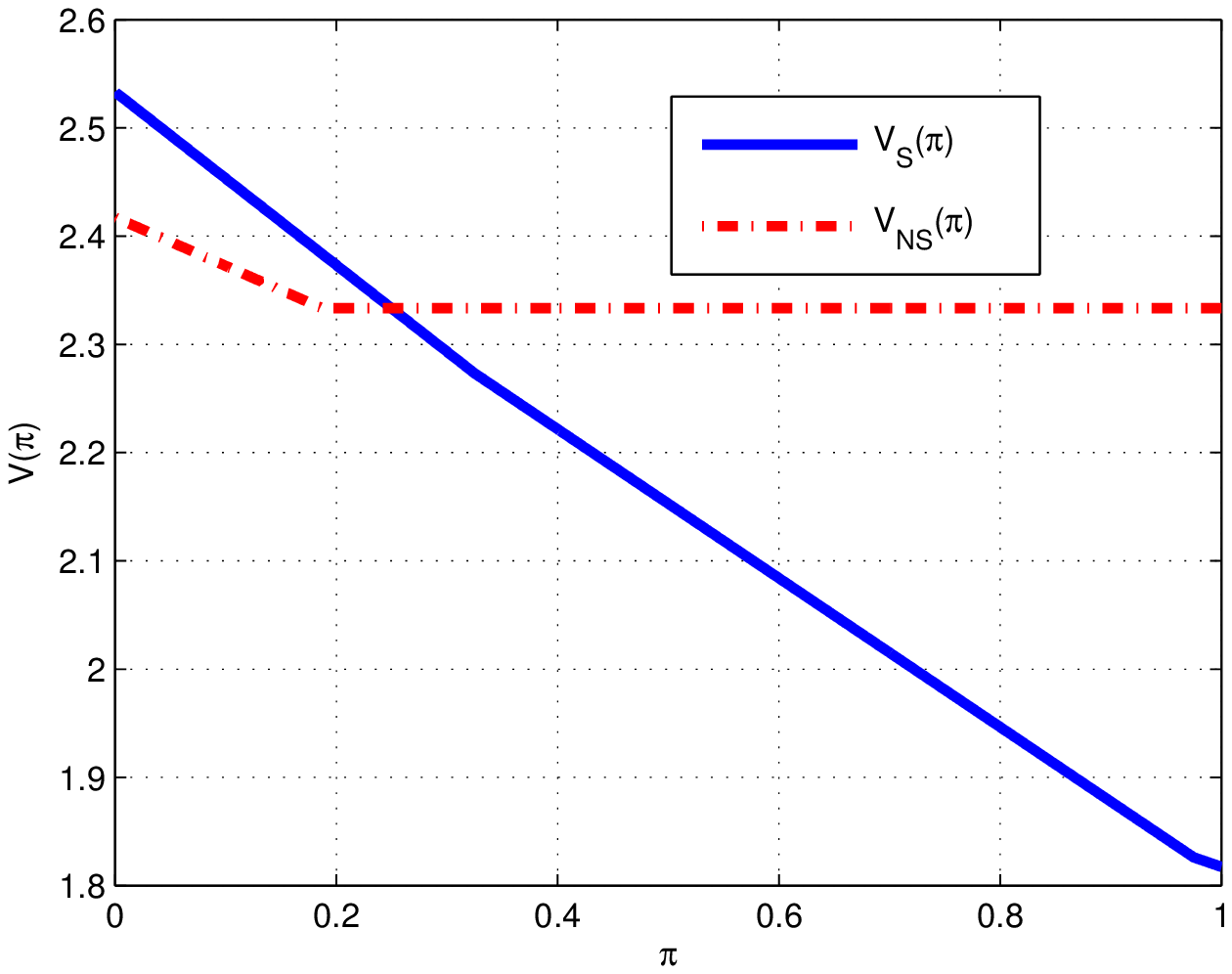} \\
     $\beta = 0.99;$ $\pi_T=0.248.$ &  $ \beta = 0.6;$ $\pi_T=0.248.$ 
    \end{tabular}

    $\eta_2=0.7;$ $\pi^{\circ} = 0.25.$
  \end{center}

  \begin{center}
    \begin{tabular}{cc}
      \includegraphics[width=0.8\columnwidth]{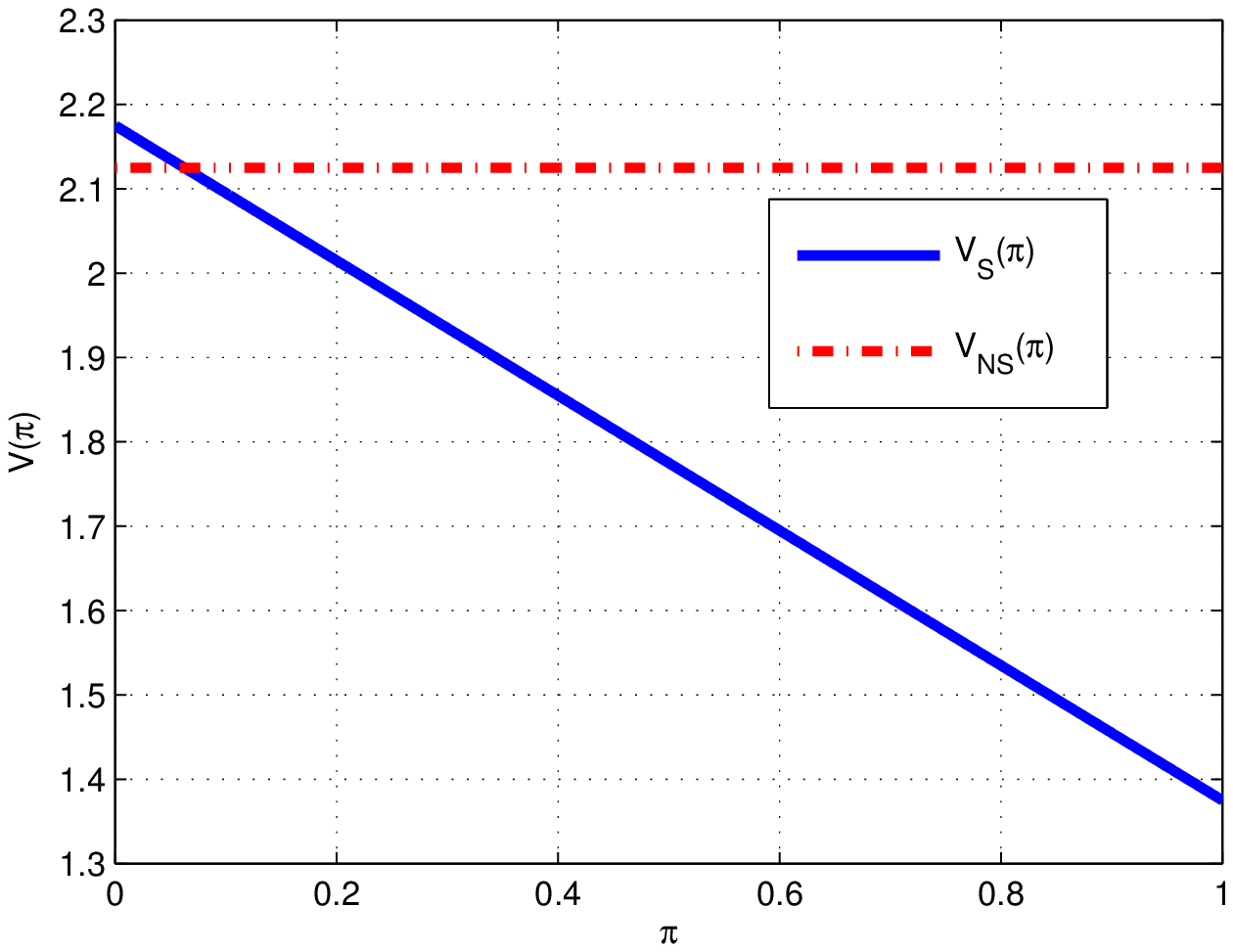}
      & 
      \includegraphics[width=0.8\columnwidth]{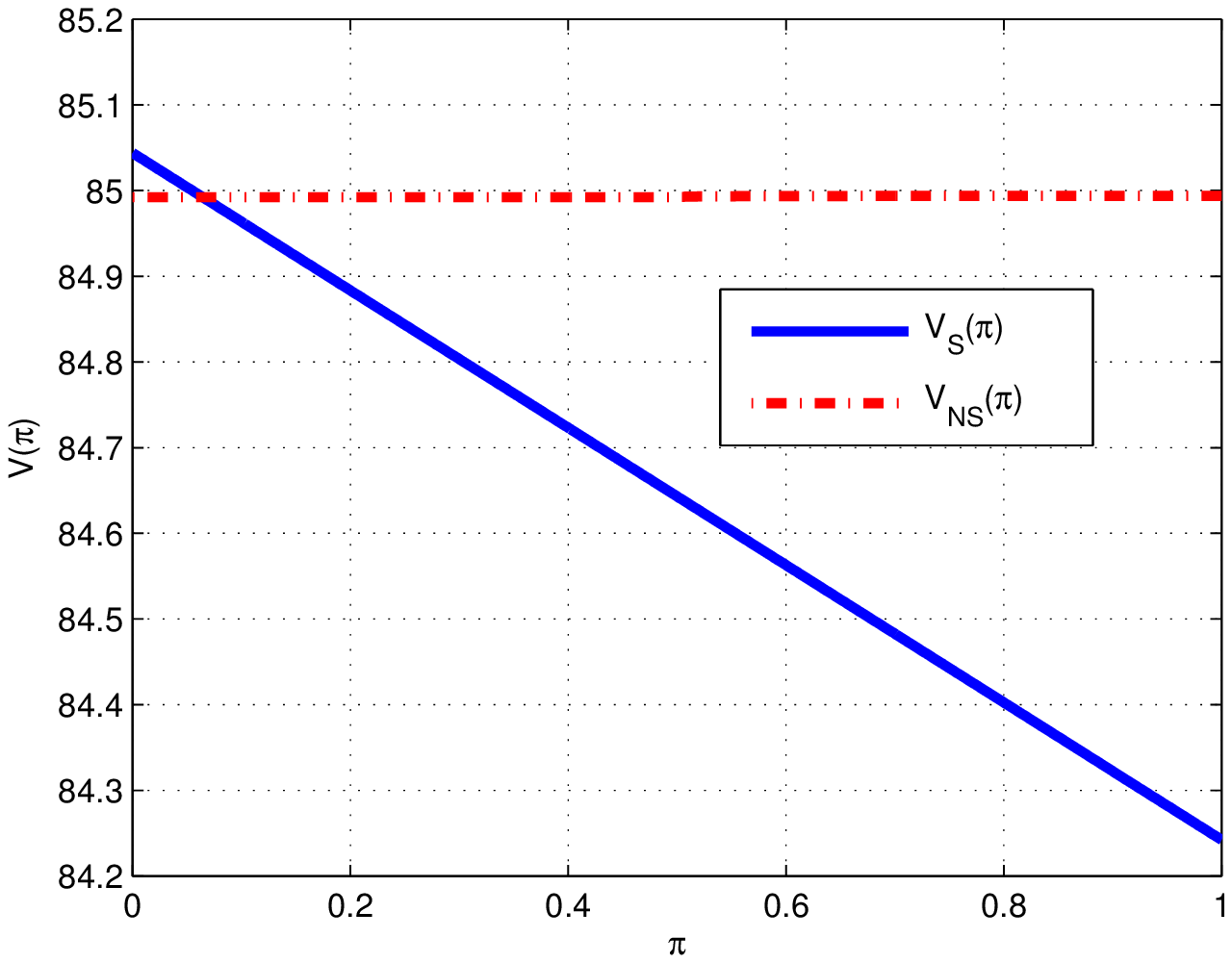} \\
      $\beta = 0.99$ $\pi_T = 0.06$&  $ \beta = 0.6;  \pi_T=0.06.$ 
    \end{tabular}

    $\eta_2=0.85;$ $\pi^{\circ} = 0.0625.$ 
  \end{center}

  \caption{$V_{NS}(\pi),$ and $V_S(\pi)$ are plotted for different
    $\eta_2$ and $\beta.$ Observe the single threshold in all the
    cases. The threshold $\pi_T$ and the $\pi^{\circ}$ are also
    indicated for each case. Here we have used $\rho_0= \eta_0 = 0.1,$
    $\rho_1=\eta_1 = 0.9,$ $\mu_0 = 0.1,$ $\mu_1 = 0.9,$ $\lambda_0 =
    0.9,$ and $\lambda_1 =0.1.$ }
  \label{plots:VS-VNS-set1}
\end{figure*}

We present some numerical results  and plot $V_{S}(\pi)$ and $V_{NS}(\pi)$ for different 
values of $\beta,$ the $\eta_2$ $\mu_i,$  $\lambda_i,$ and $\rho_i.$  The sample plots in 
Fig.~\ref{plots:VS-VNS-set1} and in many others that we computed indicate that there is
only threshold. 

\subsection{Proof of Lemma~\ref{lemma:indexability}}
\begin{IEEEproof}
  We will establish the contrapositive, i.e., assuming that $\pi_T(\eta_2)$ is {\em
    not} a monotonically decreasing function of $\eta_2$ at
  $\hat{\eta}_2$, we will show that
  \begin{equation*}
    \frac{\partial V_S(\pi,\eta_2)}{\partial \eta_2}\biggr 
    \rvert_{\pi = \pi_T(\hat{\eta}_2)} 
    \geq
    \frac{\partial V_{NS}(\pi,\eta_2)}{\partial \eta_2}\biggr 
    \rvert_{\pi = \pi_T(\hat{\eta}_2)} .
  \end{equation*}

  Suppose there exists a $\hat{\eta}_2 \in [\eta_L,\eta_H]$ such that
  $\pi_T(\eta_2)$ is increasing at $\hat{\eta}_2$ i.e., there exists a 
  $c > 0,$ such that for all $\epsilon \in [0, c ]$   
  \begin{equation*}
    \pi_T({\hat{\eta}_2}) \leq \pi_T(\hat{\eta}_2 + \epsilon).
  \end{equation*}
  This implies that for all $\epsilon \in (0,c)$
  \begin{equation}
    V_S\left(\pi_T(\hat{\eta}_2), \hat{\eta}_2 + \epsilon \right) \geq 
    V_{NS}\left(\pi_T(\hat{\eta}_2),\hat{\eta}_2 + \epsilon \right).
    \label{eq:VS-great-VNS}
  \end{equation}
  Further, from the definition of $\pi_T(\eta_2),$ we also have 
  \begin{equation}
    V_S\left(\pi_T(\hat{\eta}_2), \hat{\eta}_2\right) = 
    V_{NS}\left(\pi_T(\hat{\eta}_2),\hat{\eta}_2\right).
    \label{eq:VS-equal-VNSatr-2hat}
  \end{equation}
  Using \eqref{eq:VS-great-VNS} and \eqref{eq:VS-equal-VNSatr-2hat} we 
  can write the following.
  \begin{eqnarray*}
     &&V_S\left(\pi_T(\hat{\eta}_2),\hat{\eta}_2 + \epsilon \right) - 
     V_S\left(\pi_T(\hat{\eta}_2), \hat{\eta}_2\right)\\
     &&\geq \ V_{NS}\left(\pi_T(\hat{\eta}_2),\hat{\eta}_2 + \epsilon \right)  
     - V_{NS}\left(\pi_T(\hat{\eta}_2),\hat{\eta}_2\right).
  \end{eqnarray*}
  Dividing both sides of the above inequality by $\epsilon,$ taking limits
  as $\epsilon \to 0,$ and evaluating at $\pi = \pi_T(\hat{\eta}_2)$ gives us 
  \begin{equation*}
    \frac{\partial V_S(\pi)}{\partial \eta_2}\biggr \rvert_{\pi = \pi_T(\hat{\eta}_2)} 
    \geq
    \frac{\partial V_{NS}(\pi)}{\partial \eta_2}\biggr \rvert_{\pi = \pi_T(\hat{\eta}_2)} .
\end{equation*}
This completes the proof.
\end{IEEEproof}

\subsection{Numerical Examples}
\label{sec:rmab-numericals}

We discussed the difficulties in obtaining closed-form expression for
either of $V(\pi),$ $\pi_T(\eta_2),$ or $W(\pi)$ in some detail in
Section~\ref{sec:discuss-hidden-states}. A simple solution would be to
numerically evaluate and precompute the $W(\pi)$ by suitably
discretizing the $(0,1)$ interval. We use this technique and performed
several simulation experiments to evaluate the goodness of the
Whittle-index policy as compared to a simpler myopic policy that would
simply index the arms using $[\pi_n(t)\eta_0 + (1-\pi_n(t))\eta_1]$
for arm $n.$ This is the expected instantaneous payoff when the arm is
sampled in slot $t.$

A sample of the numerical results is presented for the following parameters 
for  a 10-armed bandit.
\begin{eqnarray*}
  \eta_0 &=&  [0.1, 0.1, 0.2, 0.4, 0.2, 0.1, 0.3, 0.3, 0.35, 0.05] \\
  \eta_1 &=& [0.9, 0.95, 0.8, 0.9, 0.6, 0.5, 0.95, 0.7, 0.85, 0.5]\\
  \mu_0 &=& [0.1, 0.9, 0.3, 0.9, 0.3, 0.9, 0.3, 0.8, 0.9, 0.5] \\
  \mu_1 &=& [0.9, 0.1, 0.9, 0.3, 0.9, 0.3, 0.9, 0.3, 0.1, 0.02]\\
  \lambda_0 &=& [0.9, 0.9, 0.1, 0.1, 0.9, 0.9, 0.9, 0.8, 0.9, 0.5] \\
  \lambda_1 &=& [0.1, 0.1, 0.8, 0.8, 0.4, 0.3, 0.4, 0.3, 0.1, 0.02].
\end{eqnarray*}
Further, $\rho_0 = \eta_0,$ and $\rho_1 = \eta_1.$ 

In the simulation, the arm with the highest index is chosen to be
played in each slot. The simulations start the arms in a random state
and a random belief about the state of the arm. In each slot one arms
is chosen to be played according to the given policy (Whittle-index
based or myopic). The reward obtained in each slot is stored and these
rewards are averaged over a $K$ iterations. The data is collected
after of 2000 slots.

\begin{figure*}

  \begin{center}
    \begin{tabular}{cc}
      \includegraphics[width=0.9\columnwidth]{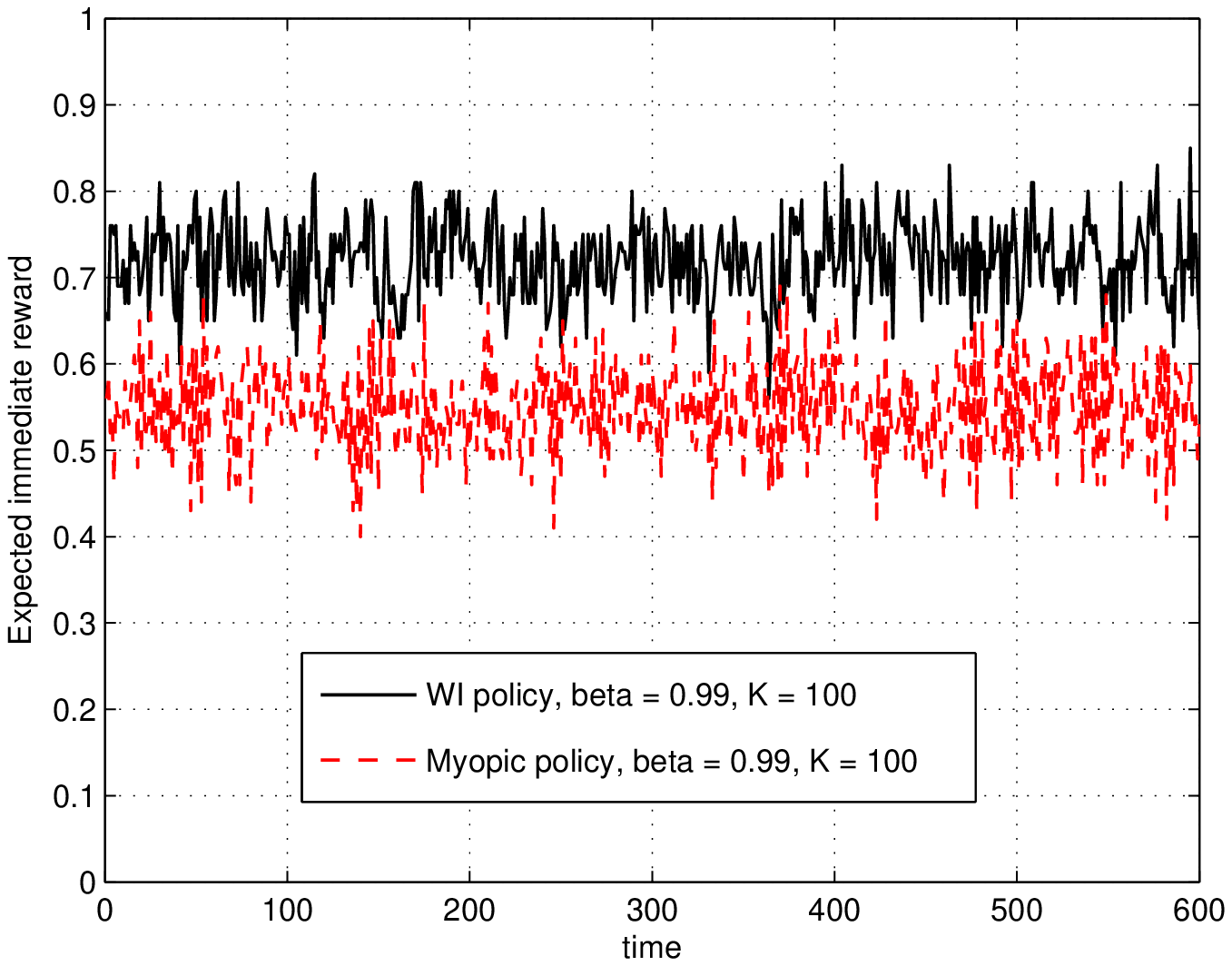}
      &
      \includegraphics[width=0.74\columnwidth]{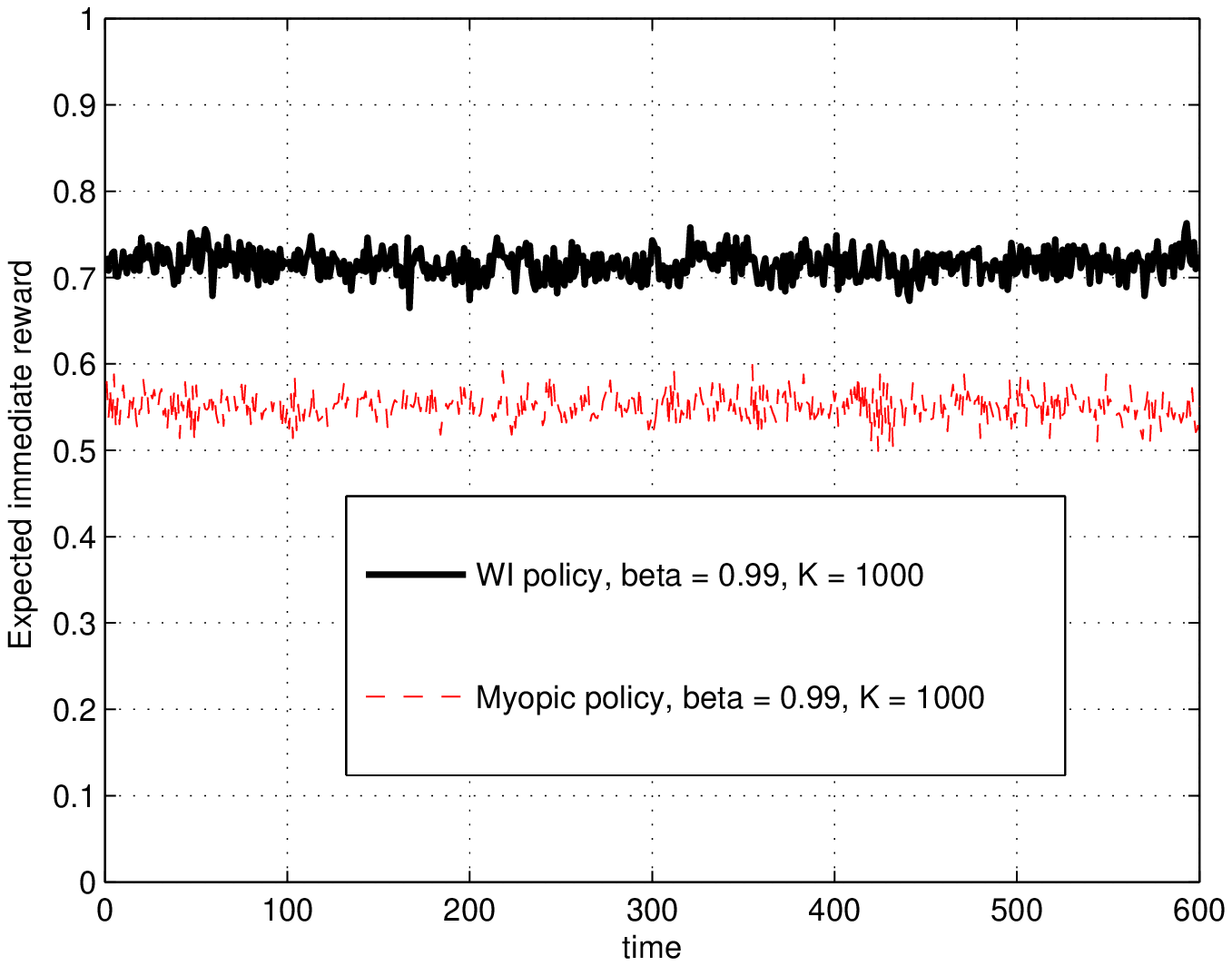} \\
      $\beta=0.99, \ K=100$ & $\beta=0.99 \ K=1000$ \\
      \includegraphics[width=0.9\columnwidth]{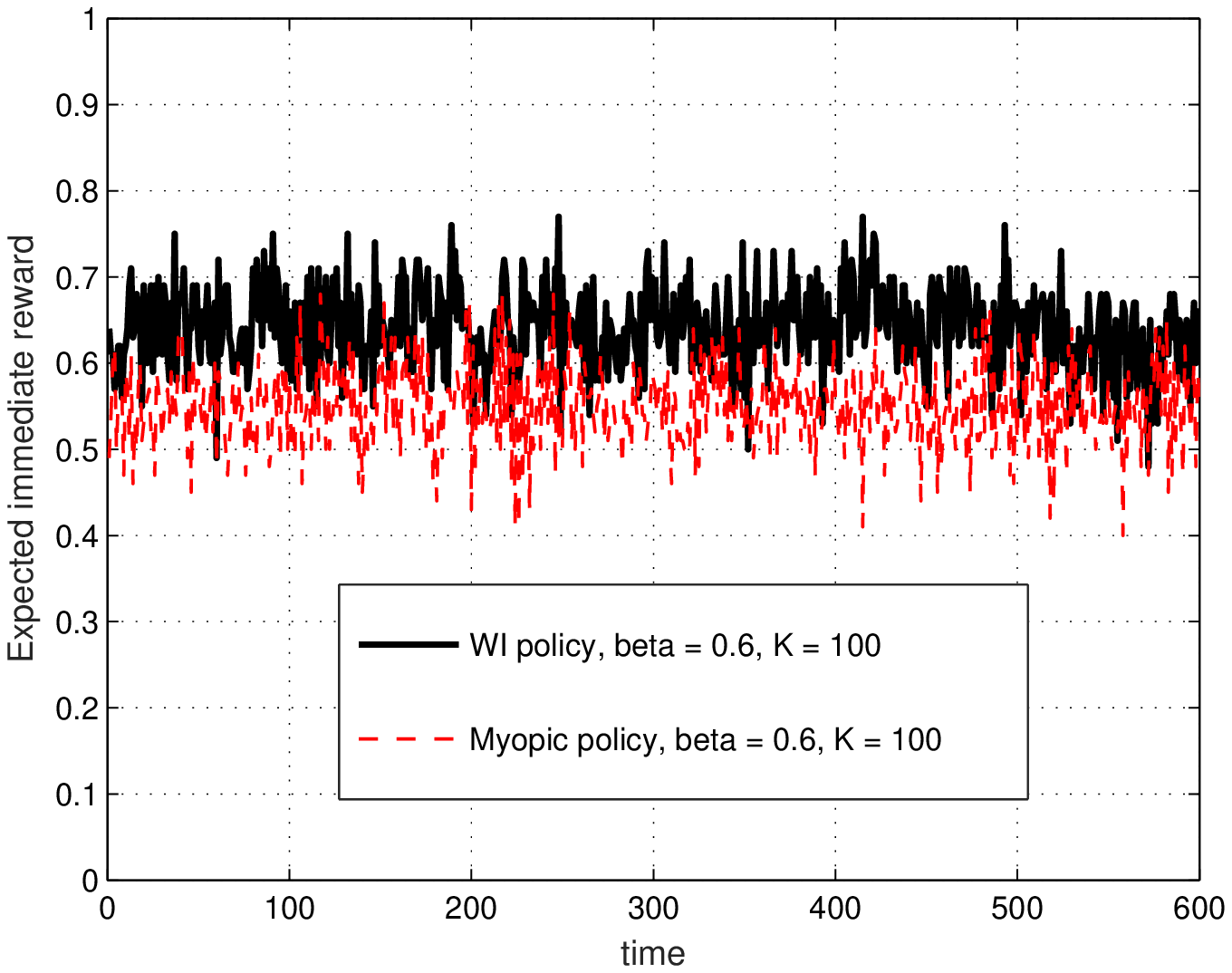}
      &
      \includegraphics[width=0.74\columnwidth]{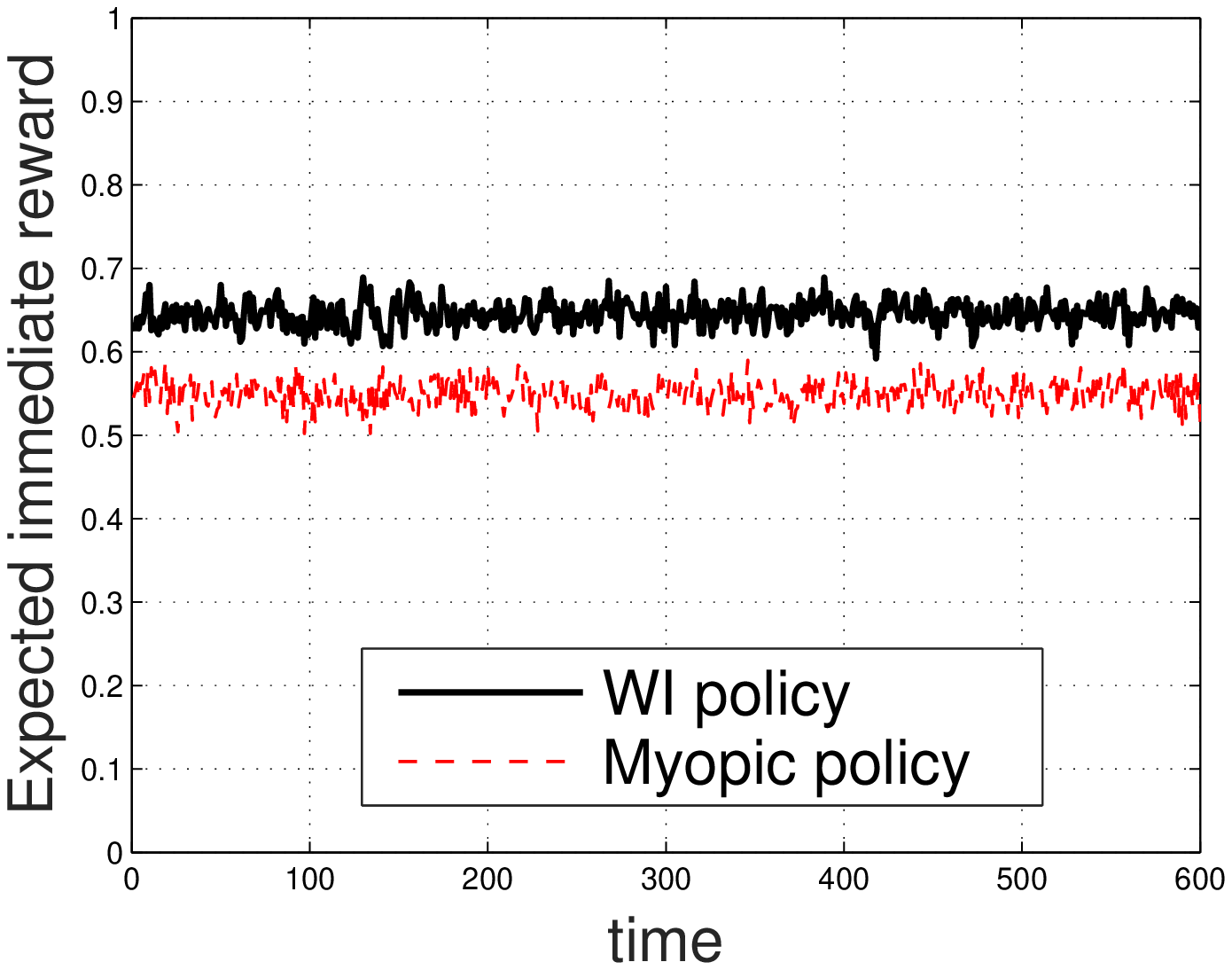} \\
      $\beta=0.6, \ K=100$ & $\beta=0.6, \ K=1000$ \\
      \includegraphics[width=0.9\columnwidth]{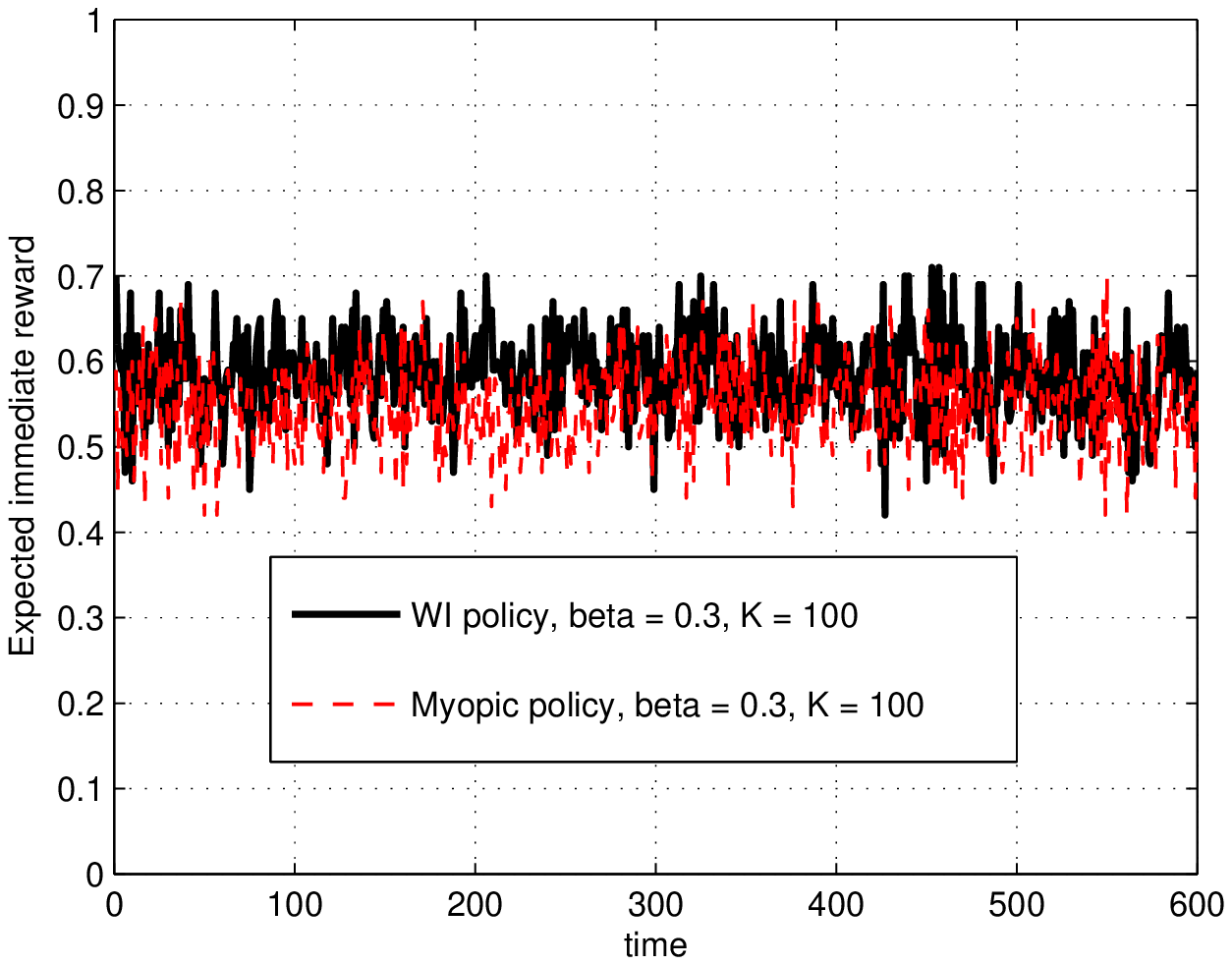}
      &
      \includegraphics[width=0.74\columnwidth]{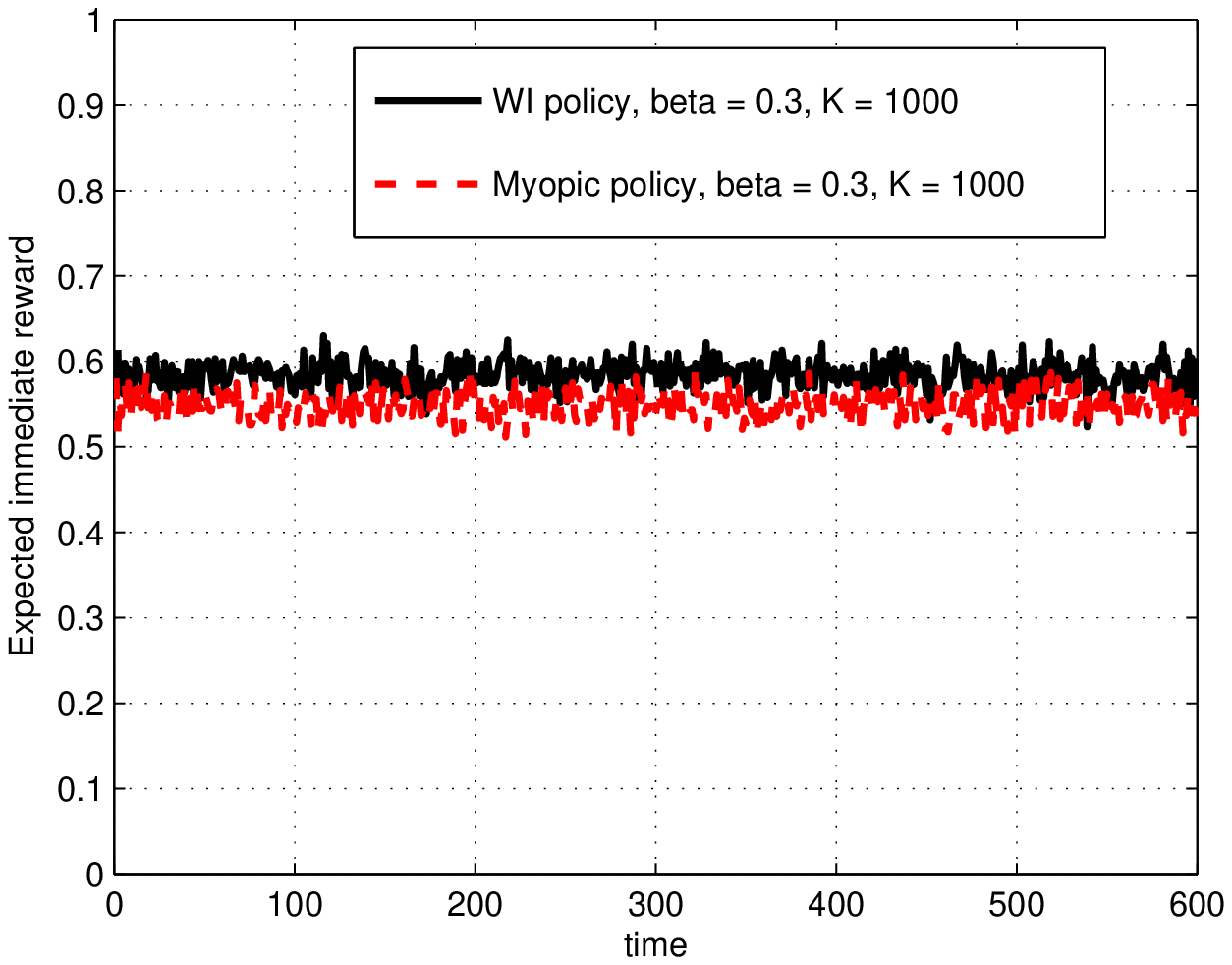} \\
      $\beta=0.3, \ K=100$ & $\beta=0.3, \ K=1000$ \\
     \end{tabular}
  \end{center}
  
  \caption{The average instantaneous reward obtained from the
    Whittle-index based policy for different values of $\beta$ and for
    the myopic policy. The average reward shown is averaged over 100
    and 1000 iterations. }
  \label{plots:Whittle-myopic}
\end{figure*}

Fig.~\ref{plots:Whittle-myopic} plots the instantaneous value of the
reward averaged over $K$ iterations for different values of $\beta$
and $K.$ For The Whittle-index policy has a consistently better reward
than than the myopic policy although the difference reduces with
decreasing $\beta.$ Our extensive simulations indicate similar
behaviour for a large set of parameters with the two becoming
comparable in a few cases.

\subsection{Complications due to hidden states}
\label{sec:discuss-hidden-states}

In this paper we are able to provide a structural property through
Theorems~1 and 2, but a obtain a closed-form expressions for the value
function $V(\pi),$ the threshold $\pi_T(\eta_2),$ or the Whittle's
index $W(\pi)$ have been elusive. We briefly discuss the complications
that the hidden states of the arms that makes it difficult to obtain
these quantities as compared to the other extant models.

Most models in the literature assume that when an arm is sampled, its
state is correctly observed. In our model, this means that when the
arm is sampled, the binary signal could just correspond to the state
of the arm and have $\rho_0=0$ and $\rho_1=1.$ In this case,
$\gamma_0(\pi) = \mu_0$ and $\gamma_1(\pi) = \mu_1$ both of which are
independent of $\pi.$ Compare this with the $\gamma_i$s for our model
that are non linear functions of $\pi!$ Further, in the models where
the state is observed, we will have
\begin{eqnarray*}
  V_S(\pi) & = & (1-\pi) + \beta (1-\pi) V(\mu_1) +\beta \pi V(\mu_0) , \\
  V_{NS}(\pi) & = & \eta_2 + \beta V(\gamma_2(\pi)) .
\end{eqnarray*}
This means that $V_S(\pi)$ can be evaluated by evaluating $V(\pi)$ at
two points.  Further, the structure of the optimal policy will be to
continue to sample while the sampled arm is observed to be in the good
state. If the arm is sampled to be in the bad state, then wait till
$\pi$ crosses $\pi_T$ before sampling again. The number of slots to
wait for this is easy to determine if $\pi_T$ is known. In our case,
if the arm is sampled and a binary $1$ is observed, the new $\pi$
depends on the current value of $\pi$ and a policy like above will not
work. A similar argument applies if the arm is sampled and a $0$ is
observed.

While obtaining closed-form expressions appears to be hard the
following properties of the $\gamma$s, obtained from first and second
derivatives, may be useful in obtaining approximations. We will not
explore that in this paper.

\begin{property}
  \label{prop:gammas}
  \begin{enumerate}
  \item If $\lambda_0 < \lambda_1$ then $\gamma_2(\pi)$ is linear
    decreasing in $\pi.$ Further, $\lambda_0 \leq \gamma_2(\pi) \leq
    \lambda_1.$
  \item If $\lambda_0 > \lambda_1$ then $\gamma_2(\pi)$ is linear
    increasing in $\pi.$ Further, $\lambda_1 \leq \gamma_2(\pi) \leq
    \lambda_0.$
  \item If $\mu_0 > \mu_1$ then $\gamma_1(\pi)$ is convex increasing
    in $\pi.$ Further, $\mu_1 \leq \gamma_1(\pi) \leq \mu_0.$
  \item If $\mu_0 > \mu_1$ then $\gamma_0(\pi)$ is concave increasing
    in $\pi.$ Further, $\mu_1 \leq \gamma_0(\pi) \leq \mu_0.$
  \item $\gamma_0(0) = \gamma_1(0) = \mu_1$ and $\gamma_0(1) =
    \gamma_1(1) = \mu_0.$ Further, if $\mu_0 > \mu_1$ then
    $\gamma_1(\pi) < \gamma_0(\pi)$ for $0 < \pi < 1.$
  \end{enumerate}
  \qed
\end{property}

\end{document}